\documentclass[11pt]{article}

\usepackage{mathdots}
\usepackage{amssymb,amsmath}
\usepackage{mathrsfs}
\usepackage{graphicx}
\usepackage[colorlinks=true, pdfstartview=FitV, linkcolor=black, citecolor=black, urlcolor=blue]{hyperref}
\usepackage[titletoc]{appendix}
\allowdisplaybreaks \numberwithin{equation}{section}
\newtheorem{thm}{Theorem}[section]
\newtheorem{prp}[thm]{Proposition}
\newtheorem{lem}[thm]{Lemma}
\newtheorem{dfn}[thm]{Definition}
\newenvironment{defn}{\begin{dfn} \rm }{\end{dfn}}
\newtheorem{cor}[thm]{Corollary}
\newtheorem{example}[thm]{Example}

\newtheorem{remark}[thm]{Remark}

\newenvironment{rmk}{\begin{remark} \rm }{\hfill $\Box$ \end{remark}}
\newenvironment{prf}{\noindent {\it Proof:} \ }{\hfill $\Box$}

\newcommand\od{\mathrm{d}}
\newcommand\ad{\mathrm{ad}}

\newcommand{\nn}{\nonumber}
\newcommand\pd{\partial}
\newcommand{\ld}{\lambda} \newcommand{\Ld}{\Lambda}
\newcommand{\al}{\alpha}
\newcommand{\gm}{\gamma} \newcommand{\Gm}{\Gamma}
\newcommand{\sg}{\sigma}
\newcommand{\om}{\omega} 
\newcommand{\ep}{\epsilon} 
\newcommand{\dt}{\delta} 
\newcommand{\ta}{\theta} \newcommand{\Ta}{\Theta}

\newcommand{\im}{\mathrm{Im}}

\newcommand\sL{\mathscr{L}} 
\newcommand\fg{\mathfrak{g}} \newcommand\cH{\mathcal{H}}
\newcommand\Z{\mathbb{Z}}
\newcommand\C{\mathbb{C}}
\newcommand\R{\mathbb{R}}
\newcommand\N{\mathbb{N}}

\newcommand{\p}{\partial}

\newcommand{\bt}{\mathbf{t}}  
\newcommand{\rs}{\mathrm{s}}

\def\Gr{\mathrm{Gr}}
\def\Id{\mathrm{Id}}

\def\bea{\begin{eqnarray}}
\def\eea{\end{eqnarray}}

\allowdisplaybreaks
\begin{document}

\title{Borodin--Okounkov formula, string equation and topological solutions of Drinfeld--Sokolov hierarchies}
\author{Mattia Cafasso$^{\dagger}$ and  Chao-Zhong Wu$^{\ddagger}$ \\
   \footnotesize $\dagger$ \footnotesize LAREMA, UMR 6093, UNIV Angers,\\
    \footnotesize CNRS, SFR Math-STIC. cafasso@math.univ-angers.fr \\
   $\ddagger$
   \footnotesize School of Mathematics, Sun Yat-Sen University \\
   \footnotesize Guangzhou 510275, P.R. China.
 wuchaozhong@sysu.edu.cn }

\date{}
\maketitle

\begin{abstract}

We give a general method to compute the expansion of topological tau
functions  for Drinfeld--Sokolov hierarchies associated to an
arbitrary untwisted affine Kac--Moody algebra. Our method consists
of two main steps: first these tau functions are expressed as
(formal) Fredholm determinants of the type appearing in the
Borodin--Okounkov formula, then the kernels for these determinants
are found using a reduced form of the string equation. A number of
explicit examples are given.
\end{abstract}

\tableofcontents

%\newpage

\section{Introduction}
Witten's conjecture \cite{Witten1}, in 1991, revealed a surprising
connection between the intersection theory of the Deligne--Mumford
moduli space $\overline{\mathcal{M}}_{g,n}$ and the Kowteweg--de
Vries (KdV) hierarchy. More precisely, let
$\langle\tau_{k_1}\tau_{k_2}\dots\tau_{k_n}\rangle_g$ be the
intersection numbers on $\overline{\mathcal{M}}_{g,n}$, and collect
them into a formal generating function (called the free energy)
\begin{equation}\label{freeenergy}
\mathcal{F}(q_0,q_1,q_2,\dots;\ep) := \sum_{g\ge0}\ep^{2
g-2}\sum_{\substack{n\ge0 \\ k_1, \ldots, k_n \geq 0}}
\langle\tau_{k_1}\tau_{k_2}\dots\tau_{k_n}\rangle_g\frac{q_{k_1}q_{k_2}\dots q_{k_n}}{n!}.
\end{equation}
Witten conjectured that this generating function is uniquely determined by these two conditions:
\begin{itemize}
    \item $U(q_0,q_1,q_2,\dots;\ep) := \ep^2 \dfrac{\partial^2 \mathcal F}{\partial q_0^2}$ satisfies the (rescaled) equations of the KdV hierarchy
        \begin{equation}\label{KdVIntro}
            \frac{\partial U}{\partial q_i} = \frac{\partial}{\partial q_0}R_i[U],
        \end{equation}
        where the differential polynomials $R_i[U]$ are defined recursively by
        \begin{equation}\label{Lenard}
            R_0[U] = U, \quad \frac{\partial R_{i+1}}{\partial q_0} = \frac{1}{2i + 3} \Big(\frac{\partial U}{\partial q_0} + 2U \frac{\partial}{\partial q_0} + \frac{\epsilon^2}4 \frac{\partial^3}{\partial q_0^3} \Big)R_i.
        \end{equation}
    \item $\mathcal{F}$ solves the string equation
        \begin{equation}\label{SEIntro}
            \sum_{k\ge1}q_k\frac{\p\mathcal{F}}{\p q_{k-1}}+\frac{q_0^2}{2\ep^2}=\frac{\p\mathcal{F}}{\p q_0}.
\end{equation}
\end{itemize}
In other words, $\tau := {\rm e}^{\mathcal{F}}$ is a tau
function %\footnote{The normalisation of the flows is a bit different from the standard one, see also below.}
of the KdV hierarchy,
uniquely selected by the string equation. Witten's conjecture has
been proved by Kontsevich \cite{Kontsevich}, and the tau function
above is usually called the Witten--Kontsevich tau function.

Since then, physicists and mathematicians working on Gromov--Witten
theory extended this beautiful connection between generating
functions of geometric invariants and integrable hierarchies to
other examples. The first (and probably the most famous) example is
the seminal work of Okounkov and Pandharipande on the Gromov--Witten
invariants of $\mathbb P^1$ and the  2D-Toda hierarchy
\cite{OKPaP1}.

The theory of Frobenius manifolds, introduced by Dubrovin in 1990s
\cite{Dub} (see also the earlier works of Saito \cite{Saito1}),
gives a general explanation of the appearance of integrable
hierarchies in Gromow--Witten theory. In particular Dubrovin and
Zhang \cite{DubZha} associated to each semi-simple Frobenius
manifold an integrable hierarchy together with a tau function,
called the \emph{topological} tau function. This topological tau
function is uniquely selected, among the solutions of the
hierarchies, by some Virasoro constraints generalizing the string
equation \eqref{SEIntro}. Moreover, the Dubrovin--Zhang topological
tau function coincides with the total descendant potential defined
by Givental \cite{GivFr2, GivTod1}, who gave a different
interpretation of Frobenius manifolds as Lagrangian cones in an
(infinite-dimensional) symplectic space. The quantization of these
Lagrangian cones leads ultimately to the extension of Gromov--Witten
theory from genus zero to all genera.

In the theory of integrable systems, Drinfeld--Sokolov hierarchies
are one of the most studied generalizations of the KdV one. These
hierarchies are indexed by affine Kac--Moody algebras, and the KdV
one corresponds to $A_1^{(1)}$. Indeed, soon after \cite{Witten1},
Witten \cite{Witten4} extended his conjecture to the general
$A_\ell$ case\footnote{Since this paper is mainly concerned with
untwisted affine Kac--Moody algebras, in the sequel we will drop the
superscript ``(1)'' when no confusion arises.} and mentioned, in
less details, a possible further generalization for the cases
$D_\ell$ and $E_\ell$. His conjecture has been proved for the
$A_\ell$ case by Faber--Shadrin--Zvonkine \cite{FSZ}, and for the
$D_\ell$ and $E_\ell$ cases by Fan--Jarvis--Ruan \cite{FJR}. Also
topological solutions of Drinfeld--Sokolov hierarchies for the non
simply-laced algebras $B_\ell,C_\ell,F_4,G_2$ are important, since
they appear in the theory of Fan, Jarvis and Ruan inspired by Witten, as developed recently by Liu, Ruan and Zhang
\cite{LRZ}. Note that for the non simply-laced cases, such
topological solutions cannot be obtained from the Dubrovin--Zhang
tau functions associated to semi-simple Frobenius manifolds
\cite{DLZ, LRZ}.

Briefly speaking, the Fan--Jarvis--Ruan--Witten (FJRW) theory is an intersection theory on the stable moduli space $\overline{\mathcal{M}}_{g,n}$ with $2 g-2+n>0$, associated to a nondegenerate quasi-homogeneous polynomial $W$ together with its symmetry group $G$.
This theory involves an $\ell$-dimensional state space $\mathscr{A}_{W, G}$ and a certain cohomological field theory $\{\Lambda_{g,n}^{W,G}\}$ (see \cite{FJR} for the construction)
of the form
\[
\Lambda_{g,n}^{W,G}: (\mathscr{A}_{W,G})^{\otimes n}\to H^*(\overline{\mathcal{M}}_{g,n}).
\]
Suppose that a basis $\{\xi_1,\xi_2,\dots,\xi_\ell\}$ of $\mathscr{A}_{W,G}$ is chosen, and  that $\psi_i:=c_1(L_i)$ is the first Chern class associated to the $i$-th tautological line bundle $L_i$ on $\overline{\mathcal{M}}_{g,n}$. For these classes the FJRW invariants, or the {genus-$g$ $n$-point correlators}, are defined by
\begin{equation}\label{FJRWinv}
\langle\xi_{i_1}\psi_1^{k_1},\cdots,\xi_{i_n}\psi_n^{k_n}\rangle_{g}^{W,G}
:=\int_{ \overline{\mathcal{M}}_{g,n} }\Lambda_{g,n}^{W,G}(\xi_{i_1},\cdots,\xi_{i_n}) \prod_{m=1}^n\psi_m^{k_m}
\end{equation}
with $i_m=1,2,\dots,\ell$ and $k_m=0,1,2,\dots$. Such invariants can be encoded in the so-called total potential function
\begin{equation}
\mathcal{F}^{W, G}= \sum_{g \geq 0} \ep^{2g-2} \mathcal{F}_g^{W,G},
\label{potential}
\end{equation}
where
\begin{equation}\label{FgW}
\mathcal{F}_g^{W,G}:= \sum_{\substack{n \geq \max\{ 0, \, 3-2g\}} } \sum_{\substack{ 1\leq i_1,\dots, i_n\leq \ell \\ k_1, \dots, k_n\ge0 }}\langle \xi_{i_1}\psi_1^{k_1}, \ldots, \xi_{i_n}\psi_n^{k_n} \rangle_{g}^{W,G} \frac{q_{i_1, k_1} \ldots q_{i_n, k_n}}{n!}
\end{equation}
with $q_{i,k}$ being formal parameters. For the total potential function, one has the following results: %In particular, the following theorem was obtained by considering the quasi-homogeneous polynomials $W$ for the ADE simple singularities together with their maximal diagonal symmetry groups, as well as their $\Gm$-reductions induced by the symmetries of simply-laced Dynkin diagrams.
%\begin{thm}[\cite{FJR, D_4, LRZ} ] \label{ADE hie}
%The following assertions are true:
\begin{itemize}
\item[(i)] As it was proved in \cite{FJR, D_4}, the total potential functions for the quasi-homogeneous polynomials $W$ associated to the $A_\ell$, $D_\ell^T$ and $E_{6,7,8}$ singularities with the maximal diagonal symmetry groups $G_{max}$ are the topological solutions of $A_\ell$, $D_\ell$ and $E_{6,7,8}$ Drinfeld--Sokolov hierarchies; the total potential functions for $D_{2 \ell} ~(\ell\ge2)$ with the symmetry groups $\langle J\rangle$ generated by $J=\left(\exp\left(\frac{2\pi\sqrt{-1}}{2 \ell-1}\right), \exp\left(\frac{2\pi\sqrt{-1}\,(\ell-1)}{2 \ell-1}\right) \right)$ are the topological solutions of the $D_{2 \ell}$ Drinfeld--Sokolov hierarchies.
\item[(ii)] As it was proved in \cite{LRZ}, the total potential functions for the $\Gm$-invariant sector of the FJRW-theory for
$D^T_{\ell+1}$, $A_{2\ell-1}$ and $E_6$ singularities with $G_{max}$ are the topological solutions of the $B_\ell$, $C_\ell$ and $F_4$ Drinfeld-Sokolov hierarchies; the total potential function for the $\Z/3\Z$-invariant sector of the FJRW-theory for
$(D_4, \langle J\rangle)$ is the topological solution of the $G_2$ Drinfeld-Sokolov hierarchy.
\end{itemize}
%\end{thm}

In view of the considerations above, it is clear that being able to
compute the expansion of topological tau functions for integrable
hierarchies is significant. Indeed, it is fair to say that this is
the main point of the Witten conjecture and its generalizations:
using the theory of integrable hierarchies to compute geometrical
invariants. For the case of KdV, and with less details for the
general $A_\ell$ case, Itzynkson and Zuber \cite{IZ}, using the
results by Kac and Schwarz \cite{KacSchwarz,SchwarzSE}, gave some
explicit expansions of the topological tau function. Besides, by now
some closed formulas for the Witten--Kontsevich tau function are
also known \cite{Al, Zh1,BDY}. In other cases, in principle one can
use Givental's quantized symplectic transformation or the axiomatic
theory of Dubrovin and Zhang to compute the topological tau
function. Nevertheless, the authors of this article are aware of
just few explicit computations besides $A_\ell$ cases (see the
previous references and also \cite{ LVX, LRZ, LYZ, Zh2}).

The main objective of this paper is to give a general and completely
algorithmic way to compute the topological tau function of
Drinfeld--Sokolov hierarchies associated to an arbitrary untwisted
affine Kac--Moody algebra. To some extent, our results are a
far-reaching generalization of Itzynkson and Zuber's results, as we
also use the Sato--Segal--Wilson theory \cite{SW} together with
(properly defined) Kac--Schwarz operators associated to affine
Kac--Moody algebras. Moreover, we make a systematic use of the
identification between tau functions and the Szeg\H{o}--Widom constant
for large (block) Toeplitz determinants, building on our previous
results obtained in \cite{C1, CafassoWu1}. Indeed, among these tau
functions, a crucial step is to select the one satisfying the string
equation, which implies the Virasoro constraints as studied by one
of the authors in \cite{WuVir} (cf. also \cite{HMSG}).

Our main objective is achieved in Theorem \ref{mainthm}. Using this
theorem, one can easily re-derive (with few lines of code) the
expansions contained in section 5. Our expansions agree with the one
in \cite{Zh2} for the $A_2$ case, and for the cases $D_4, B_3, C_2$
with the recent results obtained with a different method in
\cite{LRZ}. It should be mentioned that, in our setting, one does
not even need to write down explicitly the nonlinear evolutionary
equations defining the hierarchy, but only certain data of the
affine Kac--Moody algebra. Moreover, we prove that the topological
tau functions in the BCFG cases can be reduced from those in the ADE
cases, which agrees with the main result of Liu--Ruan--Zhang
\cite{LRZ}.

The structure of the paper is as follows:
\begin{itemize}
    \item The second section is dedicated to the connection between block Toeplitz
    determinants and tau functions. It extends to the case of formal series the results
    obtained in \cite{C1,CafassoWu1}. The main results of the section are
    Theorems \ref{thmtaufunction} and \ref{thm-tauHH}. In particular the latter, connecting
    the Borodin--Okounkov formula with the Baker-Akhiezer function of the hierarchy,
    establishes a relation between topological tau functions and discrete integrable
    operators \cite{BorodinOkounkov, BorodinDiscretePainleve}.%, which we plan to study elsewhere.
    \item The third section is built on the results obtained in \cite{WuVir}.
    Our aim is to find the point in the Grassmannian associated to the topological tau function for an arbitrary Drinfeld--Sokolov hierarchy. To this end, we reduce the string equation to a very simple form  on the affine Kac--Moody algebra, and prove that the reduced equation has a unique solution that determines the point in the Grassmannian. We also show how  to extract, in the $A_\ell$ case, the well known Kac-Schwarz operators \cite{KacSchwarz} from the reduced string equation. In this sense, the results of this Section can be considered as a generalization of \cite{KacSchwarz} to the case of an arbitrary untwisted affine Kac--Moody algebra.
    %, which can be solved uniquely on the  hence generalizing the results obtained by Kac and Schwarz in \cite{KacSchwarz} for the $A_\ell$ case. Some details on the relations between this section and \cite{KacSchwarz} are contained in the Appendix A.
    \item In the fourth section, we state and prove our main result, Theorem \ref{mainthm}, and study reduction properties
    of the topological tau functions.
    \item The fifth section contains the first terms of the expansions of the topological tau
    functions for $A_\ell$ $(\ell = 1,2,3)$, $D_4$, $B_3$ and $C_2$, computed using our
    algorithm. The results are consistent with the existing
    literature.
%\item The last section is inspired by the
%    latest work \cite{BDY} of Bertola, Dubrovin and Yang, who obtained generating
%    functions for multi-point correlation functions of the KdV hierarchy. In
%    this appendix, we will derive similar generating functions for the cases $A_{\ell}$, $B_{\ell}$, $C_{\ell}$,
%    $D_{\ell}$ and $G_2$, which is an alternative to compute the
%    correlators $\langle\tau_{k_1}\tau_{k_2}\dots\tau_{k_n}\rangle_g$ by using the solutions of the
%     reduced string equation for the Drinfeld--Sokolov hierarchies.
    \item In Appendix A, for the convenience of the reader, we add a list of
    matrix realizations for the Kac--Moody algebras used in this paper. This list
    is extracted from \cite{DS, Kac}. \\

\end{itemize}

\section{Grassmannians and Toeplitz operators}\label{sec-SSW}

The aim of this section is to extend the results of \cite{CafassoWu1} to the realm of formal series.
We start recalling the so-called Borodin--Okounkov
formula for the case of (formal) block Toeplitz determinants.

\subsection{The Borodin--Okounkov formula}

We want to consider formal power series of the form
$$\varphi(z) = \sum_{k \in \mathbb Z} \varphi_k z^k, \quad\quad \varphi_k = \sum_{j \geq 0} \varphi_{k,j} \lambda^j, \quad\quad \varphi_{k,j} \in \mathfrak{gl}(n).$$
We will call such series \emph{formal loops}. We will need to be able to multiply them and, in order to do so, we introduce a gradation with respect to $\lambda$.
\begin{defn}
For any $f = \sum_{j \geq 0} f_j \lambda^j \in \mathbb{C}[[\lambda]]$, its $\lambda$--degree (denoted with $\deg_\lambda f$) is defined as the smallest $j$ such that $f_j \neq 0$. If $f \equiv 0$, by definition $\deg_\lambda f = + \infty.$
\end{defn}
\begin{defn}
	Given a series $\Big(f^{(k)} \Big)_{k = 0}^\infty$ of elements in $\mathbb C[[\lambda]]$, we say that $f \in \mathbb C[[\lambda]]$ is the limit of the series (i.e., $\lim_{k \rightarrow \infty} f^{(k)} = f$) if $\lim_{k \rightarrow \infty} \deg_\lambda(f^{(k)} - f) = + \infty.$
\end{defn}
Now we fix a positive integer $h$ and we define \emph{$h$-admissible} (formal) loops.
\begin{defn}\label{equivassumption}
Let $\varphi = \sum \varphi_k z^k$ be a loop. We say that the loop is $h$-admissible if there exists a positive integer $h$ such that, for every integer $k$ such that $|k| \geq 2$,
\begin{equation}\label{assu}
          \deg_\ld(\varphi_{k})_{ij} \geq  (|k|-1) h \quad \forall i, j = 1\ldots n.
\end{equation}
\end{defn}
When there is no ambiguity, we will speak about admissible loops, without mentioning the integer $h$.
At this stage, Definition \ref{equivassumption} seems artificial but it is natural
from the point  of view of affine Kac--Moody algebras, as it will be
clearer in the next sections. %In particular, in the scalar case, our
%condition is a generalization of the one in \cite{BorodinOkounkov}.
Note that, given two admissible loops, their product is well defined (but possibly not $h$-admissilbe).\\

We also define the space of vector-valued formal power series
$$H^{(n)} := \Big\lbrace v(z) = \sum_{k \in \Z} v_k z^k\; | \, v_k \in \C^n[[\lambda]] \Big\rbrace,$$
For any admissible loop $\varphi$ the operator of multiplication
$$\varphi : H^{(n)} \longrightarrow H^{(n)}$$
is well defined.\\
The vector space $H^{(n)}$ is spanned by the standard
vectors $\{z^k{\rm e}_\al\mid \al=1,\ldots,n; k\in\Z\}$, where ${{\rm
e}_{\alpha}}$ is the column vector with its $\al$-th component being $1$
and the other components vanish.\\
With respect to this basis, any vector
$v(z)=\sum_k v_k z^k \in H^{(n)}$ can be identified with its
coordinates as
\begin{equation}\label{convention}
    v \sim \left(\begin{array}{c}\vdots \\ v_{-1} \\ v_0 \\ v_1 \\ \vdots
    \end{array}\right).
\end{equation}
It is not hard to see that, with respect to this basis, the operator of multiplication by an admissible loop has a (block) matrix representation given by the \emph{Laurent} matrix
$$L(\varphi) := (\varphi_{s-t})_{s,t\in \mathbb Z}.$$
In the sequel, given an admissible loop $\varphi$ we will be interested in the following associated $\mathbb{N} \times \mathbb{N}$ matrices, where $\mathbb N$ denotes the set of non--negative integers:
\begin{equation}\label{ToeplitzHankel}
    T(\varphi) := \Big(\varphi_{s-t}\Big)_{s,t \in \N}; \quad H(\varphi) := \Big(\varphi_{s+t+1}\Big)_{s,t \in \N}; \quad \widetilde H(\varphi) := \Big(\varphi_{-s-t-1}\Big)_{s,t \in \N}.
\end{equation}
The first matrix is the so-called (block) Toeplitz matrix associated
to $\varphi$, while the second and the third are the two associated
Hankel matrices. It is common to call $\varphi$ the \emph{symbol} of such matrices.\\
Let $H^{(n)}_-$ and $H^{(n)}_+$ be the subspaces generated
respectively by the negative and non--negative Fourier modes, so
that as usual we have $H^{(n)} = H^{(n)}_- \oplus H^{(n)}_+$. We
will denote with $p_\pm$ the projections onto the corresponding
subspaces. We also introduce the following involution operator
\begin{align}\label{}
\iota: H^{(n)} &\rightarrow H^{(n)} \nn\\
v(z) & \mapsto  v(z^{-1})z^{-1}.
\end{align}
Clearly, $\iota\circ \iota=\Id$, and the restrictions $\iota:
H^{(n)}_\pm \rightarrow H^{(n)}_\mp$ are one-to-one correspondences. Then, one can verify that the matrices in \eqref{ToeplitzHankel} are the matrix representation of the following endomorphisms of $H_+^{(n)}$, well defined for any admissible loop :
\begin{equation}\label{oprepresentation}
    T(\varphi) = p_+ \circ \varphi{|_{H_+^{(n)}} }; \quad H(\varphi) = p_+ \circ \varphi \circ \iota{|_{H_+^{(n)}} }; \quad \widetilde H(\varphi) = \iota\circ p_- \circ \varphi{|_{H_+^{(n)}} }.
\end{equation}

The following lemma is well known (see, for instance,
\cite{BotSilb}):
\begin{lem}\label{lemmahaenkeltoeplitz}
    Given two admissible loops $\varphi_1,\varphi_2$, we have the following identity between $\N \times \N$ matrices:
    \begin{equation}\label{toeplitzhaenkelidentity}
        T(\varphi_1)T(\varphi_2) = T(\varphi_1\varphi_2) - H(\varphi_1)\widetilde H(\varphi_2).
    \end{equation}
 In particular, $T(\varphi_1)T(\varphi_2) = T(\varphi_1\varphi_2)$, whenever $T(\varphi_1)$ is block upper-triangular or $T(\varphi_2)$ is block lower-triangular.
\end{lem}
\begin{prf}
 Starting from the left hand side, we have
 \begin{eqnarray}\nonumber
    T(\varphi_1)T(\varphi_2) &=& p_+ \circ \varphi_1 \circ p_+ \circ \varphi_2|_{H_+^{(n)} }
    = p_+ \circ \varphi_1  \circ (\Id - p_-) \circ \varphi_2|_{H_+^{(n)}}
    \\ &=& p_+ \circ (\varphi_1\varphi_2){|_{H_+^{(n)}} } -
    p_+\circ\varphi_1\circ \iota\circ \iota \circ p_- \circ \varphi_2|_{H_+^{(n)} } \nn\\
    &=& T(\varphi_1\varphi_2) - H(\varphi_1)\widetilde H(\varphi_2).
 \end{eqnarray}
 Thus the lemma is proved.
\end{prf}

It is worth noticing that, when a symbol $\varphi$ is admissible, the entries in the $(s,t)$-block of the Hankel matrices
$H(\varphi)$ and $\widetilde H(\varphi)$ have a $\lambda$-degree greater than
$(|s+t+1|-1)h$. In particular, both of them are
Hilbert--Schmidt operators, in the sense that the norm
$$|H(\varphi)|^2_{HS} := \sum_{i,j\geq 0} |(H(\varphi))_{i,j}|^2$$
is well defined as the limit of the truncated sums (and the same for $\widetilde H(\varphi)$).

Given a positive integer $N$, let  $T_N(\varphi)$ denote the
$(N+1)\times(N+1)$ upper-left principal minor of $T(\varphi)$, that
is,
\[
T_N(\varphi):= \left(\begin{array}{cccc}
\varphi_0 & \varphi_{-1} & \ldots & \varphi_{-N}\\
&&&\\
\varphi_1 & \varphi_0 & \ldots & \varphi_{-N+1}\\
&&&\\
   \vdots       & \vdots & \ddots & \vdots\\
&&&\\
\varphi_N & \varphi_{N-1} & \ldots & \varphi_0
\end{array}\right),
\]
and its determinant is denoted as $D_N(\varphi)$. What is more, let
us denote with $P_N : H^{(n)}_+ \longrightarrow H^{(n)}_+$ the
projection to the space of formal series of type $\sum_{0 \leq k
\leq N} v_kz^k$, and $Q_N := \Id - P_N$. We are ready to writing
down the celebrated Borodin--Okounkov formula, as it appears in
\cite{BasorWidomToeplitz}.
\begin{thm}[Borodin--Okounkov formula \cite{BorodinOkounkov}]\label{BOthm}
Consider an admissible loop $\varphi$ and suppose that it admits two distincts factorizations $$\varphi = \varphi_+\varphi_- = \psi_-\psi_+,$$
where all the terms are admissible loops, $\varphi_+,\psi_+$ are power series in $z$ and $\varphi_-,\psi_-$ in $z^{-1}$. Suppose moreover that the zero Fourier modes of $\psi_-,\varphi_-$ are equal to the identity and the zero Fourier modes of $\psi_+,\varphi_+$ are equal to the identity plus strictly lower triangular matrices.
Then, for any positive integer $N$,
$$D_N(\varphi) = Z(\varphi) \det(\Id - K_N),$$
where $Z(\varphi) := \det\Big(\Id - H(\varphi)\widetilde H(\varphi^{-1})\Big)$, $K := H(\varphi_-\psi_+^{-1})\widetilde H(\psi_-^{-1}\varphi_+)$
and $K_N := Q_N K Q_N$.
\end{thm}
\begin{remark}\label{traceclassremark}
 Note that both the determinants written above are well defined, since they differ
 from the identity by a trace-class operators (recall that the product of two Hilbert-Schmidt operators is of trace class).
\end{remark}
\begin{prf}
For the readers' convenience, we adapt here the proof given by E.
Basor and H. Widom in \cite{BasorWidomToeplitz}, writing it directly for the
matrix case, adapted to the setting of formal series. Our aim is to
compute the determinant of $T_N(\varphi)$, which is the upper--left
block of size $(N+1)n\times (N+1)n$ of the matrix $P_N T(\varphi)
P_N$. Note that we will always work with the standard basis for
$H^{(n)}_+$ described above.

Using Lemma \ref{lemmahaenkeltoeplitz} and the two (obviously
satisfied) equations
$$P_N T(\varphi_+) = P_N T(\varphi_+) P_N,\quad T(\varphi_-) P_N= P_N T(\varphi_-) P_N$$
we have the following chain of equalities
\begin{eqnarray*}
    P_N T(\varphi) P_N &=& P_N T(\varphi_+) T(\varphi_+^{-1})T(\varphi)T(\varphi_-^{-1})T(\varphi_-) P_N\\
                    &=& P_N T(\varphi_+)P_N T(\varphi_+^{-1})T(\varphi)T(\varphi_-^{-1})P_N T(\varphi_-) P_N.
\end{eqnarray*}
Note that the determinants of the upper-left block of $P_N
T(\varphi_\pm) P_N$ are unity because of the form of the symbols.
This leads to the fact that $D_N(\varphi)$ %the determinant of $T_N(\varphi)$
is nothing but the determinant of the upper--left block of $P_N M
P_N$,  where $$M := T(\varphi_+^{-1})T(\varphi)T(\varphi_-^{-1}).$$
On the other hand, we can immediately observe that $M$ is similar,
via the invertible operator $T(\varphi_+)$, to the matrix
$T(\varphi)T(\varphi^{-1}) = \Id - H(\varphi)\widetilde
H(\varphi^{-1})$, whose (formal) Fredholm determinant is well
defined, as observed before (and so is the determinant of $M$).
Moreover, $M$ is invertible and the inverse is
\begin{equation}\label{inverseA}
    M^{-1} = T(\varphi_-)T(\psi_+^{-1})T(\psi_-^{-1})T(\varphi_+),
\end{equation}
as one can check directly, again using Lemma \ref{lemmahaenkeltoeplitz}. Hence
\begin{eqnarray*}
    D_N(\varphi) &=& \det(T_N(\varphi)) = \det(P_NMP_N + Q_N) \\
            &=& \det(M(P_N + M^{-1}Q_N)(\Id-Q_NMP_N))\\
         &=& \det(M)\det(P_N + M^{-1}Q_N)\det(\Id - Q_NMP_N).
\end{eqnarray*}
We have already checked $\det(M) = Z(\varphi)$. Moreover $\det(\Id -
Q_NMP_N) = 1$ because $Q_NP_N = 0$. Hence it remains to prove
$$\det(P_N + M^{-1}Q_N) = \det(\Id - K_N).$$
This is finally done using the chain of equalities
%\begin{eqnarray*}
    $$\det(P_N + M^{-1}Q_N) = \det(\Id- (\Id - M^{-1})Q_N) = \det(\Id - Q_N(\Id- M^{-1})Q_N)$$
%\end{eqnarray*)
and equation \eqref{inverseA}.
\end{prf}
\begin{remark}\label{remdeg}
    Suppose that one wants to compute the Szeg\"o--Widom constant $Z(\varphi)$ for an
    admissible symbol $\varphi$ up to a certain degree. Using the Borodin--Okounkov formula, one has
    $$\deg_\ld(\log Z(\varphi) - \log D_N(\varphi)) = \deg_\ld(\log \det (\Id - K_N)),$$
    Since each entry of $K_N$ has $\ld$ degree greater or equal to $Nh$, we have
    $$\deg_\ld(\log Z(\varphi) - \log D_N(\varphi)) \geq Nh.$$
\end{remark}
A slightly better estimation will be proved later for a particular class of symbols we are interested in, see Proposition \ref{comp}.

In the limit for large $N$ we obtain a (formal) version of the Szeg\H{o}--Widom theorem.
\begin{cor}
  Under the conditions of Theorem \ref{BOthm}
  $$\lim_{N\rightarrow \infty} D_N(\varphi) = Z(\varphi) = \det\Big(\Id - H(\varphi)\widetilde H(\varphi^{-1})\Big).$$
\end{cor}

\subsection{Sato's Grassmannian and the related tau function}

To each admissible loop of the form
$$\gamma(z) = \Id + \sum_{k < 0} \gamma_k z^k$$ we associate the subspace
$$W_\gamma := \gamma \cdot H^{(n)}_+$$ and we denote with $\Gr^0$ the set of all the subspaces
$W \subseteq H^{(n)}$
obtained in this way. This is the so--called big cell of the Sato--Segal--Wilson Grassmannian. Observe that, for each $W \in \Gr^{(0)}$, the projection $p_+\,:\,W\longrightarrow H^{(n)}_+$ is  an isomorphism and $z W\subseteq W$.
%Let the Grassmannian $\Gr^0$ be the set of all subspaces $W
%\subseteq H^{(n)}$ such that the projection
%$p_+\,:\,W\longrightarrow H^{(n)}_+$ is  an isomorphism and $z W
%\subseteq W$. To every space $W \subseteq H^{(n)}$, one can
%associate a (formal) loop $\gamma$ % with value in $GL_n(\mathbb{C})$
%such that
%\[
%W = W_\gamma := \gamma H^{(n)}_+, \quad \gamma = \Id +
%\mathcal{O}(z^{-1}).
%\]

Denote with $G_+$ the group of admissible loops of the form
$$g(z) = \Id + \sum_{k \geq 0} g_k z^k$$ and consider an (arbitrary) abelian
subgroup $G^a_+ \subseteq G_+$. We proceed to recall the definition
of the Sato--Segal--Wilson tau function for a point in the
Grassmannian $\Gr^0$ acted upon by an abelian group $G_+^a$.

Given $W_\gamma \in\Gr^0$, we consider $\gamma$ as a map from
$H_+^{(n)}$ to $H^{(n)}$ by left multiplication.
%Then the point $W=W_\gamma$
%is described via its $\Z\times\N$ matrix representation as
%\begin{equation}\label{Wwpm}
%W\sim  %\left(\begin{array}{c}
%                    \om_-\\
%                    \om_+
%                    \end{array}\right) =
%\Big(\gamma_{j-k}\Big)_{j \in \Z, k \in \N}:=
%\left(\begin{array}{cccc}
%\vdots &\vdots &\vdots & \iddots  \\
%\gm_{-1}&\gm_{-2}&\gm_{-3}&\dots \\
%\gm_0& \gm_{-1}&\gm_{-2}&\dots \\
%\gm_1& \gm_0& \gm_{-1}&\dots \\
%\vdots &\vdots &\vdots &  \ddots \\
  %  \end{array}\right),
%\end{equation}
%with blocks being the Fourier coefficients of $\gamma$.
Let $\om_\pm := p_\pm\circ\gamma$, these being maps
\[
\om_\pm: H_+^{(n)}\longrightarrow  H_\pm^{(n)}.
\]
We also introduce an operator
\begin{equation}\label{}
h_{W_\gamma}:H^{(n)}_+\to H^{(n)}_-
\end{equation}
whose graph is $W_\gamma$, namely,
\begin{equation}\label{}
h_{W_\gamma}:=\om_-\circ \om_+^{-1}=p_-|_{W_\gamma}\circ(p_+|_{W_\gamma})^{-1}
\end{equation}
(we have used the property that $p_+|_{W_\gamma}$ is one-to-one).\\
Every element $g\in G_+^a$ defines by multiplication a map
\[
g: H^{(n)} \longrightarrow H^{(n)}.
\]
Its inverse can be written in matrix form as
$$g^{-1}=\left(\begin{array}{ccccc}
                    d & 0\\
                    b & a
                    \end{array}\right)$$
where
\[
a:H^{(n)}_+\to H^{(n)}_+,\quad b: H^{(n)}_-\to H^{(n)}_+,\quad d:
H^{(n)}_-\to H^{(n)}_-.
\]
\begin{dfn}[\cite{SW}]\label{def-SWtau}
Given a point $W_\gamma\in \Gr^0$, the associated Sato--Segal--Wilson (SSW) tau function
depending on $g\in G_+^a$ is defined as
\begin{equation}\label{tauSSW}
\tau_{W_\gamma}(g) := \det\Big(\Id+a^{-1} \circ b \circ h_{W_\gamma} \Big) = \det\Big(\Id+b \circ h_{W_\gamma}\circ a^{-1}\Big).
\end{equation}
\end{dfn}

The careful reader might observe that, since we are working on the
setting  of formal series, it is not clear if the determinants above
are well defined. Nevertheless, thanks to the Theorem \ref{thmtaufunction} below, the determinants in \eqref{tauSSW} are identified with the Szeg\H{o}--Widom constant of an admissible loop, and hence well defined because of Remark \ref{traceclassremark}.

Given a point $W_\gamma \in \Gr^{0}$, define
\begin{equation}
    J_\gamma(z) := g^{-1}(z)\gamma(z),
\end{equation}
where $g \in G_+^a$ (and then, in particular, it is admissible). Then, the following theorem holds
true
\begin{thm}\label{thmtaufunction}
    For any point $W_\gamma \in \Gr^0$ we have
    $$\tau_{W_\gamma}(g) = Z(J_\gamma) = \det (\Id - H(J_\gamma)\widetilde H(J_\gamma^{-1})),$$
    where the equality is understood as an equality of formal series in $\ld$.
\end{thm}
\begin{prf}
The proof for the case of formal series is the same as the one in
the analytical  case discussed in \cite{CafassoWu1}. We report it
here solely for the readers' convenience. By the definition of the
SSW tau function we have
\[
\tau_{W_\gamma}(g) = \det(\Id + b\circ\omega_-\circ\omega_+^{-1}\circ
a^{-1})
\]
where the terms in the determinant can be written as follows:
\begin{eqnarray}
    &a^{-1} = p_+ \circ g|_{H^{(n)}_+};& b = p_+ \circ g^{-1}|_{H^{(n)}_-} ;\\
    &\omega_+^{-1} = p_+ \circ \gamma^{-1}|_{H^{(n)}_+};& \omega_- = p_- \circ \gamma{|_{H_+^{(n)}} }.
\end{eqnarray}
Hence we obtain (here and below we suppress the sign of
composition)
\begin{equation}\label{eqn1}
    \tau_{W_\gamma} = \det\Big(\Id + p_+  g^{-1}  p_-  \gm \;  p_+  \gm^{-1} g|_{H_+^{(n)} }\Big).
\end{equation}
On the other hand, using the Szeg\H o--Widom theorem, we have
\begin{eqnarray}
    Z(J_\gamma) &=& \det\Big(\Id - H(J_\gamma)\widetilde H(J_\gamma^{-1})\Big)
     = \det\Big(\Id - p_+  g^{-1}  \gamma \; \iota \; \iota \; p_-  \gamma^{-1} g{|_{H_+^{(n)}} }\Big)\nonumber \\
    &=& \det\Big(\Id - p_+  g^{-1}  \gamma \; p_-  \gamma^{-1} g{|_{H_+^{(n)}} }\Big). \label{eqn2}
\end{eqnarray}
Combining \eqref{eqn1} and \eqref{eqn2}, it is sufficient to prove
\begin{equation}\label{eqn3}
    \left.\left(p_+  g^{-1}  p_-  \gm \; p_+  \gm^{-1}  g + p_+  g^{-1}
    \gamma \; p_-  \gamma^{-1} g\right)\right|_{H_+^{(n)}}  = 0.
\end{equation}
Indeed, the left hand side is
\begin{align}\label{}
\mathrm{l.h.s.}=& \left.\left(-p_+  g^{-1}  p_+  \gm \; p_+ \gm^{-1}
g + p_+  g^{-1}  \gm \; p_+ \gm^{-1} g + p_+  g^{-1}
    \gamma \; p_-  \gamma^{-1} g\right)\right|_{H_+^{(n)}} \nn\\
=& \left.\left(-p_+  g^{-1} \om_+  \; \om_+^{-1} g + p_+ g^{-1} \gm
\;  \gm^{-1} g \right)\right|_{H_+^{(n)}}
=\left.\left(-\Id + \Id\right)\right|_{H_+^{(n)}}=0.
\end{align}
Thus we conclude the theorem.
\end{prf}

\subsection{The wave function and finite-size Toeplitz determinants}

Given a point $W_\gamma = \gamma H^{(n)}_+$ in the Grassmannian
acted by $g\in G^{a}_+$, we recall that we defined $J_\gamma = g^{-1}\gamma$. The
Borodin--Okounkov formula acquires a particular nice form in the
case of the (block) Toeplitz determinants associated to our symbols
with two factorizations:
\begin{equation}\label{RHJ}
    J_\gamma = g^{-1}\gamma = J^-_\gamma J^+_\gamma,
\end{equation}
where $J^-_\gamma(z) = \Id + \mathcal O(z^{-1})$, $J^+(z) = \Id +
N_- + \mathcal O(z)$, $N_-$ strictly lower triangular. To start, let
us remember the definition of Baker--Akhiezer function associated to
$W_\gamma$.
%There, we say that a matrix-valued function belongs to a certain point $W \in \Gr^{0}$ if all the columns do.

\begin{dfn}\label{def-bf}
Suppose that an element $W_\gamma\in \Gr^0$ is given. A matrix
function $w(g;z)$, depending on $g\in G_+^a$ and $z\in S^1$, is
called the Baker--Akhiezer (wave) function associated to $W_\gamma$
if
\begin{enumerate}
        \item $w(g;z)\in W_\gamma$ for all $g\in G_+^a$ (a matrix-valued function belongs to a
certain point $W \in \Gr^{0}$ if all its columns do);
        \item $p_+(g^{-1}w(g;z)) = \Id$.
\end{enumerate}
\end{dfn}
The following lemma was proved, for instance, in \cite{CafassoWu1}, where also the unicity of $w(g;z)$ is discussed.
\begin{lem}\label{wavefunctionlemma}
    Given $W_\gamma \in \Gr^{0}$, the corresponding wave function $w = w(g;z)$ is expressed as
    \begin{equation}
        w = g J_\gamma^-.
    \end{equation}
\end{lem}
\begin{prf}
    The first property defining $w$ is satisfied because
    \begin{eqnarray*}
        \gamma^{-1}w = \gamma^{-1}gJ_\gamma^- = J_\gamma^{-1}J_\gamma^- = \big(J_\gamma^+\big)^{-1} \in H_+^{(n)},
    \end{eqnarray*}
    so that $\gamma^{-1}w \in H_+^{(n)}$ and, consequently, $w \in \gamma H_+^{(n)} = W_\gamma$.\\
    The second condition is immediate, since
    $$p_+(g^{-1}w) = p_+ (J_\gamma^{-}) = \Id.$$
\end{prf}

The theorem below expresses the Borodin--Okounkov operator $K_N$ for
the symbol  $J_\gamma$ in terms of the wave function associated to
the point $W_\gamma$. Note that in a less general setting this
result was already proved in \cite{C1}.
\begin{thm}\label{thm-tauHH}
    Given a point $W_\gamma := \gamma H_+^{(n)} \in \Gr^{0}$, let
    $J_\gamma := g^{-1}\gamma$. Then
    \begin{equation}\label{maineq2}
        D_N(J_\gamma) = \tau_{W_\gamma}(g)\det(\Id - K_N)
    \end{equation}
    where $K_N = Q_N K Q_N$ with
    \begin{equation}\label{maineq3}
        K = H(w(g))\widetilde H(w^{-1}(g)).
    \end{equation}
\end{thm}
\begin{prf}
   Equation \eqref{maineq2} is simply a restating of the Borodin--Okounkov formula
   applied to the symbol $J_\gamma$, together with the explicit form \eqref{RHJ} of the two
   Riemann--Hilbert factorizations of $J_\gamma$,
   Lemma \ref{wavefunctionlemma} and Theorem \ref{thmtaufunction}.
\end{prf}
\begin{rmk}
    Theorem \ref{thmtaufunction}
    expresses the tau function as the determinant of an operator whose resolvent, $K$, is
    written in function of the corresponding wave function. In
    particular, it can be shown that
    \begin{equation}
      \tau_{W_\gamma}(g) = \det \Big(\Id - H(w(g))\widetilde H(w^{-1}(g))\Big)^{-1}.
    \end{equation}
    As already noted in \cite{BorodinOkounkov}, one interesting feature of the operator
    $K$ is that, in many cases, it turns out to be a \emph{discrete integrable} operator,
    in the sense of \cite{BorDiscrBes}. Namely, its non-diagonal elements can be written
    in the form
    \begin{equation}\label{discrker}
        K_{i,j} = \dfrac{\sum_{k} p_k(i)q_k(j)}{i - j}, \quad i\neq j.
    \end{equation}

    For the case of topological solutions of Drinfeld--Sokolov hierarchies, we
    can prove that the associated operators $K$ is indeed a discrete integrable operator.
    However, in our case the sum in \eqref{discrker}
    runs over $\N$, which makes the corresponding (discrete) Riemann--Hilbert problem
    operator-valued, as in \cite{ItsKoz}.
    The integrability is due to the fact that, because of the string equation \eqref{steTa}, the wave function satisfies the equation
    $\p_z w(z) = w(z)M(z)$, with $M(z)$ of a particular form.

     It would be interesting to see if the integrability leads to some discrete equations of
     Painlev\'e type, as for instance in
     \cite{BorodinDiscretePainleve}. We will study it elsewhere.
\end{rmk}

%\subsection{Some ingredients for explicit computations}

%Instead of computing directly the block--Toeplitz determinants, as we did till now, maybe it is easier to compute the determinants of the operators of type $\Id - H(J_\gamma)\widetilde H(J_\gamma^{-1}).$ To start, denote with $G$ the operator corresponding to this operator, i.e.
%$$\Id - H(J_\gamma)\widetilde H(J_\gamma^{-1}) = \Id - G.$$
%As you can easily check, you can write the generating functions of the entries of $G(i,j)$ as follows
%\begin{equation}
%    \sum_{i,j\geq 1} \zeta^i\eta^{-j}G(i,j) = \frac{J_\gamma(\zeta)J^{-1}_\gamma(\eta)}{1 - \eta/\zeta} %= \frac{g^{-1}(\zeta)\gamma(\zeta)\gamma^{-1}(\eta) g(\eta)}{1- \eta/\zeta}
%\end{equation}
%where the equation above should be understood expanding the denominator in the geometric power series %$\sum_{k\geq 0} \frac{\eta^k}{\zeta^k}.$ Then, we have two possibilities:

%\begin{itemize}
%   \item Either we use the standard formula
%        $$\det(\Id - G) = \sum_{m = 0}^\infty (-1)^m \sum_{1\leq \ell_1 < \ldots \ell_m}^\infty \det\Big[G(\ell_i,\ell_j)\Big]_{i,j = 1}^m.$$
%    \item Either we use the formula
%       $$\ln\det(\Id - G) = -\sum_n \frac{1}n \mathrm{Tr} G^n.$$
%\end{itemize}
%This last one, in my opinion, looks promising since we can``break out" the computations we were doing till now to more elementary bricks that, hopefully, will be easier to compute for the computer. Also, maybe that $\gamma$, in our cases, is symmetric so that we can easily write the inverse?

\section{String equation for Drinfeld--Sokolov hierarchies}

In this section we find the initial condition for the topological
solutions of Drinfeld--Sokolov hierarchies (i.e. the ones satisfying
the string equation). We will focus on the case of untwisted affine
Kac--Moody algebras.

\subsection{Review of affine Kac--Moody algebras}
\label{sec-g}

We briefly recall how to construct untwisted affine Kac--Moody
algebras \cite{Kac}. Consider a simple Lie algebra $\mathring{\fg}$
of rank $\ell$, and let $\mathring{A}:=(a_{i j})_{1\le i, j\le
\ell}$ be its Cartan matrix. Then the associated Kac--Moody algebra
$\fg$ is realized as  the central extension of the associated loop
algebra, that is
\begin{equation}\label{}
\fg=\left(\mathring{\fg}\otimes\C[z,z^{-1}]\right)\oplus\C c,
\end{equation}
and the Lie bracket is defined by
\begin{align}\label{XYbr}
&[X z^{j}+x c, Y z^k+y c]=[X, Y]z^{j+k}+\dt_{j,-k}{j}(X\mid Y)_0\,c
\end{align}
with $X, Y\in\mathring{\fg}$ and $(\cdot\mid\cdot)_0$ being the
standard invariant symmetric bilinear form on $\mathring{\fg}$. We
denote with $A$ the Cartan matrix associated to $\fg$, which is
obtained from $\mathring{A}$ by adding one column and one row:
$A=(a_{i j})_{0\le i, j\le \ell}$. The lowest positive integer
solutions $\{k_i\}_{i = 0}^\ell$ of $\sum_{j=0}^\ell a_{i j}k_j=0$
are called the Kac labels of $\fg$.

Assume that the simple Lie algebra $\mathring{\fg}$ has the
following root space decomposition
\[
\mathring{\fg}=\mathring{\mathfrak{h}}\bigoplus_{\al\in\Delta}\mathring{\fg}_\al,
\]
where $\Delta$ is the set of roots. Let $\Pi=\{\al_1, \al_2,
\dots,\al_\ell\}\subseteq \Delta$ be the set of simple roots. A
system of Weyl generators $\lbrace E_i, F_i, H_i \rbrace$  for
$\mathring{\fg}$ can be chosen in such a way that
$E_i\in\mathring{\fg}_{\al_i}$, $F_i\in\mathring{\fg}_{-\al_i}$ and
$H_i=[E_i,F_i]$. The highest root for $\mathring{\fg}$ is
$\ta=\sum_{i=1}^\ell k_i \al_i$, and one can choose
$F_0\in\mathring{\fg}_\ta$, $E_0\in\mathring{\fg}_{-\ta}$ and
$H_0=[E_0,F_0]$. It is known that the following elements
\begin{equation}\label{weyl}
e_i := z^{\dt_{i,0}}E_i, \quad f_i := z^{-\dt_{i,0}}F_i, \quad
\al^\vee_i := H_i+\dt_{i,0}c \qquad  (0\le i\le \ell)
\end{equation}
compose a set of Weyl generators for the affine algebra $\fg$.

Every non-vanishing integer vector $\rs=(s_0, s_1, \dots,
s_\ell)\in\Z^{\ell+1}_{\ge0}$ induces a gradation on $\fg$ by
\[
\deg e_i=s_i, \quad \deg f_i=-s_i,\quad  \deg \al^\vee_i=0, \qquad
0\le i\le \ell.
\]
The decomposition of spaces with respect to $\rs$ is written as
\[
\fg=\bigoplus_{k\in\Z}\fg_{k\,[\rs]},
\]
and we will use the notations $\fg_{\ge m\,[\rs]} = \sum_{k\ge
m}\fg_{k\,[\rs]}$, $\fg_{< m\,[\rs]} = \sum_{k< m}\fg_{k\,[\rs]}$,
etc. In particular, the following vectors
\begin{equation}\label{}
\rs^0=(1,0,\dots,0), \quad \rs^1=(1,1,\dots,1)
\end{equation}
give the homogeneous and the principal gradations
respectively.

Let $\mathring{\mathscr{E}}=\{m_i\}_{i=1}^\ell$ be the set of exponents of $\mathring{\fg}$, with
\[
1= m_1< m_2\le m_3\le \dots \le m_{\ell-1}
< m_{\ell}=h-1\]
and $h:=\sum_{i=0}^{\ell}k_i$ being the Coxeter number for $\fg$,
then the set $\mathscr{E}$ of the exponents of $\fg$ is
\[
\mathscr{E}=\mathring{\mathscr{E}}+h\Z.
\]
There are elements
$\Ld_j\in\fg_{j\,[\rs^1]}$ for $j\in \mathscr{E}$ such that
\begin{equation}\label{}
[\Ld_j, \Ld_k]=j\dt_{j,-k}c, \quad j, k\in \mathscr{E}.
\end{equation}
These elements generate the principal Heisenberg subalgebra
$\mathcal{H}$ of $\fg$. In particular, noting that $1$ is always an
exponent, one has
\[
\Ld_1=\nu\Ld,
\]
where $\Ld:=\sum_{i=0}^\ell e_i$ and $\nu$ is a normalization
constant (for the purpose of considering Virasoro constraints
\cite{WuVir}). It is known that $\Ld$ is a semisimple element,
namely,
\begin{equation}\label{dec2}
\fg=\cH+\im\,\ad_{\Ld}, \quad \cH\cap\im\,\ad_{\Ld}=\C c.
\end{equation}
Note $\mathrm{ker}\,\ad_\Ld=\cH$ modulo the center $\C c$.

\begin{rmk}
The construction of a twisted affine Kac--Moody algebra of type
$X_N^{(r)}$, with $r=2,3$, is more complicated. In brief, one can
choose in $\mathring{\fg}$ a set of elements $E_i$, $F_i$ and $H_i$
that are invariant with respect to a certain diagram isomorphism of
order $r$, then a system of Weyl generators of $\fg$ is given by
\eqref{weyl}, see Chapter~8 of \cite{Kac}.
\end{rmk}

In the sequel we will take an $n\times n$ trace-less matrix
realization of $\mathring{\fg}$, and get into the homogeneous
realization of $\fg$ as in the appendix (see \cite{DS, Kac}).

\subsection{Tau functions of Drinfeld--Sokolov hierarchies}

Let us recall the definition of Drinfeld--Sokolov hierarchies
\cite{DS}. Given an affine Kac--Moody algebra $\fg$, introduce an
operator
\begin{equation}\label{sL}
\sL := \frac{\od}{\od x}+\Ld+q,
\end{equation}
where $q$ is a function of $x\in\R$ taking values in the Borel subalgebra generated by $\{f_i, \al_i^\vee\mid i=1,2,\dots,\ell\}$. One sees that on the set of such kind of operators there is a class of gauge transformations defined by
\begin{equation}\label{gauge}
\sL\mapsto \tilde{\sL}= e^{\ad_X}\sL,
\end{equation}
with $X$ being a function of $x$ that takes values in the nilpotent subalgebra generated by   $\{f_i\mid i=1,2,\dots,\ell\}$.
By using the property \eqref{dec2}, one has the following
proposition.
\begin{prp}[\cite{WuVir}]\label{thm-dr0}
Given an operator $\sL$ of the form \eqref{sL}, there exists a unique
function $U$ taking values in $\fg_{<0\,[\rs^1]}$ such that the following two
conditions are fulfilled:
\begin{align}
(i)\quad   &\tilde{\sL}:=e^{-\ad_{U}}\sL=\frac{\od}{\od x}+\Ld+H \quad \hbox{with $H$ taking values in }~ \mathcal{H}_{<0\,[\rs^1]}, \label{UL}\\
(ii)\quad & \left(e^{\ad_{U}}\Ld_j\right)_c=0\quad
\hbox{ for any positive exponent }~ j\in \mathscr{E}_{>0}, \label{ULdc}
\end{align}
where the subscript ``$c$'' means to take the coordinate along $c$ with
respect to the following decomposition of $\fg$:
\[
\fg=\C\al_1^\vee\oplus\dots\oplus\C\al_\ell^\vee\oplus\C\,c
\oplus\left(\bigoplus_{k\ne0}\fg_{k\,[\rs^1]}\right).
\]
Moreover, both $U$ and $H$ are differential polynomials in (the components of) $q$.
\end{prp}

The Drinfeld--Sokolov hierarchy associated to $\fg$ is defined by the following set of
partial differential equations
\begin{equation}\label{Lt2}
\frac{\pd \sL}{\pd t_j}=[-(e^{\ad_U}\Ld_j)_{\ge0\,[\rs^0]}, \sL], \quad
j\in \mathscr{E}_{>0}
\end{equation}
restricted to an equivalence class of $\sL$ with respect to the
gauge transformations \eqref{gauge}. Here the subscript ``${\ge0\,[\rs^0]}$'' means the projection to $\fg_{\ge0\,[\rs^0]}$ with respect to the decomposition $\fg=\fg_{\ge0\,[\rs^0]}\oplus\fg_{<0\,[\rs^0]}$.  We refer the readers to \cite{DS} for more details in choosing such a gauge slice, which can be viewed as a manifold
of dimension $\ell$. We remark that, in the Drinfeld--Sokolov hierarchy one has
$\p/\p t_1=\nu\p/\p x$ due to the fact $\Ld_1=\nu\Ld$; henceforth we identify
$t_1$ with $ x/\nu$.

In \cite{HM, Mi, WuVir} a special gauge slice was chosen such that $\sL$ can be represented as
\begin{equation}\label{LTa}
\sL= \Theta\left(\frac{\od}{\od x} + \Lambda\right)\Theta^{-1} +
\omega \cdot c
%=\Ta\left(\frac{\od}{\od x}+\Ld\right)\Ta^{-1}
\end{equation}
with $\Theta$ lying in the Lie group of $\fg_{<0\,[\rs^0]}$ and
$\omega$ a scalar function (in fact $\omega=\p_x\log\tau$ with
the tau function $\tau$ given below), such that the Drinfeld--Sokolov hierarchy \eqref{Lt2} can be
written equivalently as
\begin{equation}\label{Tat}
\frac{\pd\Ta}{\pd t_j}=(\Ta\Ld_j\Ta^{-1})_{<0\,[\rs^0]}\,\Ta, \quad
j\in \mathscr{E}_{>0}.
\end{equation}
In fact, one can introduce
\begin{equation}\label{omj}
\om_j:=(\Ta \Ld_j\Ta^{-1})_c, \quad j\in \mathscr{E}_{>0},
\end{equation}
and they satisfy
\[
\pd_{t_i}\om_j=\pd_{t_j}\om_i, \quad i,j\in \mathscr{E}_{>0}.
\]
Based on this fact,
the tau function of Drinfeld--Sokolov hierarchy was introduced in
\cite{WuVir} (cf. \cite{HM, Mi} and references therein) as follows.
\begin{dfn}%[\cite{WuVir}]
\label{WuDefinition} Given a solution $\Ta$ of the Drinfeld--Sokolov
hierarchy \eqref{Tat} (or equivalently $q$ of \eqref{Lt2}), the tau
function $\tau (\bt)$, with $\bt$ being the set of the parameters
$\{t_j \mid j \in \mathscr{E}_{>0}\}$, is defined by
\begin{equation}\label{tauDS}
\pd_{t_j}\log\tau =-(\Ta \Ld_j\Ta^{-1})_c, \quad j\in \mathscr{E}_{>0}.
\end{equation}
\end{dfn}
\begin{rmk}\label{rmk-Taom}
From \cite{WuVir} one also sees that the components of $\Ta$ can be represented as polynomials in the ring
\[
\C\left[ {\pd_{t_1}}^k\om_i,\ \om_j\mid i\in\mathring{\mathscr{E}}, \  k\ge1, \  j\in \mathscr{E}_{>0}\right].
\]
\end{rmk}

%Observe that the tau function $\tau $ is determined up to a constant factor.

%\vskip 2ex

Given the $n\times n$ matrix realization of $\mathring{\fg}$, we
establish a connection with the $n$-dimensional vector valued space
$H^{(n)}$ in the previous section. In fact, as we proved in
\cite{CafassoWu1}, choosing appropriately $G_+^a$ and $\gamma$, the
tau functions $\tau$ and the Sato--Segal--Wilson $\tau_{W_\gamma}$ are
(essentially) the same. Here the equality will be established after rescaling $\Theta$ by $\lambda$.
We recall that $\Theta$ lies in the Lie group of $\fg_{<0\,[\rs^0]}$. Moreover, recalling the homogeneous realization of the Kac--Moody algebra $\fg$, the
parameter $z$ corresponds to principal degree $h$, while a constant
matrix has principal degree between $-h$ and $h$ (here we extend the principal degree of the Lie algebra to the associated matrix algebra). Note that
$\fg_{<0\, [\rs^0]}\subset \fg_{<0\, [\rs^1]}$,
hence we can write
\begin{equation}\label{principalTheta}
	\Theta(\bt ; z) = {\exp}\left( \sum_{k<0} Y_k(\bt)\right), \quad Y_k(\bt) \in \fg_{k\, [\rs^1]}
\end{equation}
and rescale it as
\begin{equation}\label{rescaledprincipalTheta}
	\tilde\Theta(\bt ; z) = {\exp}\left(\sum_{k<0} \lambda^{|k|} Y_k(\bt)\right).
%, \quad Y_k(\bt) \in \fg_{k\, [\rs^1]}
\end{equation}

Afterward, we choose as our loop $\gamma$ defining the point $W_\gamma \in \Gr^{0}$
value as
\begin{equation}\label{gm}
\gm(z) := \left.\tilde\Ta^{-1}(\bt;z)\right|_{\bt=0}.
\end{equation}
Remark that, in consideration of the form of $\tilde\Theta$, one has
$\gm(z)=I+\mathcal{O}(z^{-1})$ as $z\to\infty$.
As for $G_+^a$ we choose it to be the following:
\begin{equation}\label{Gaplus}
G_+^a=\left\{g(\bt ; z)=\exp\left(\sum_{j\in \mathscr{E}_{>0}}\lambda^j t_j
\Ld_j\right)\right\}.
\end{equation}
In particular, we have \begin{equation}\label{gmep}
\left.\begin{array}{l}
    \deg_\ld(\gm_{-k})_{ij}\\
  %  \\
    %\deg_\ld(\gm^{-1}_{-k})_{ij}\\
    \\
    \deg_\ld(g_{k}(\bt))_{ij}
\end{array}\right\} \ge (k-1)h, \quad \forall\; 1\leq i,j\leq n, \quad k\ge 2.
\end{equation}
so that both $\gamma$ and $g(\bt;z)$ are $h$--admissible loops.\\

The result below is proved in \cite{CafassoWu1} in the setting of
smooth functions (see Theorem~4.4) but there is no difficulty to
extend it to formal series.
\begin{thm}\label{thm-tautauDS}
Given a ($\lambda$--rescaled) solution of the Drinfeld--Sokolov hierarchy \eqref{Lt2}
associated to an untwisted affine Kac--Moody algebra $\fg$,
the two tau functions $\tau (\bt)$ and $\tau_{W_\gm}(\bt)$ mentioned above  % defined by \eqref{tauDS} and by \eqref{satoDS}
satisfy %(up to a constant)
\begin{equation}\label{tautauDS}
\log\tau ={\kappa}\log\tau_{W_\gm},
\end{equation}
where $\kappa$ is a constant such that $(X\mid Y)_0=\kappa\,
\mathrm{Tr}(X Y)$ gives the standard invariant bilinear form on
$\mathring{\fg}$ (indeed $\kappa=1$ for the cases of types A and C,
while $\kappa=1/2$ for types B and D, as realized in the appendix).
%the corresponding simple Lie algebra such that a long root has square length $2$.
\end{thm}

%\begin{rmk}
%More exactly, one has $\kappa=1$ for the special linear/sympletic
%algebras (types A and C) and $\kappa=1/2$ for the special orthogonal
%algebras (types B and D), see the appendix of \cite{DS}.
%%Hence $\tau=\tau_{SSW}$ in the type A case.
%\end{rmk}

Combining Theorems~\ref{thm-tautauDS} and \ref{thm-tauHH}, we
conclude  that the tau function of the Drinfeld--Sokolov hierarchy
with (rescaled) initial data $\gm(z)$ in \eqref{gm} is given by
\begin{equation}\label{tauDSHH}
\log\tau ={\kappa}\log Z(J_\gamma) = {\kappa}\log\det (\Id -
H(J_\gamma)\widetilde H(J_\gamma^{-1})),
\end{equation}
where
\begin{equation}\label{}
J_\gamma(\bt;z)=g^{-1}(\bt;z)\gamma(z).
%, \quad \gm(z)=\Ta^{-1}(\bt;z)|_{\bt=0}.
\end{equation}
This result will be applied below to compute the so-called
topological solution of the hierarchy.

\begin{rmk}
The equality \eqref{tautauDS} holds true for an arbitrary affine
Kac--Moody algebra of type $X_N^{(r)}$, with the constant factor
$\kappa$ replaced by $\kappa/(r k_0)$.
\end{rmk}

%and assumes that it has the following formal Fourier expansion
%\begin{equation}\label{}
%J_\gamma(\bt;z)=\sum_{i\in\Z}J_i z^i.
%\end{equation}
%%\begin{thm}[\cite{CW}]\label{thm-DM}
%The Sato--Segal--Wilson tau function for $W_\gamma$ coincides with
%the corresponding large-size limit of the block Toeplitz determinant
%$D_\infty(J_\gamma(\bt;z))$, that is,
%\begin{equation}\label{}
%\tau_{W_\gm}(\bt)=D_\infty(J_\gamma(\bt;z)):=\lim_{N\to\infty}\det
%T_N(J_\gamma(\bt;z))
%\end{equation}
%with
%\[
%T_N(J_\gm):= \left(\begin{array}{cccc}
%J_0 & J_{-1} & \ldots & J_{-N}\\
%&&&\\
%J_1 & J_0 & \ldots & J_{-N+1}\\
%&&&\\
%   \vdots       & \vdots & \ddots & \vdots\\
%&&&\\
%J_N & J_{N-1} & \ldots & J_0
%\end{array}\right).
%\]
%\end{thm}

\subsection{The reduced string equation}

For the Drinfeld--Sokolov hierarchy  \eqref{Lt2} associated to an
untwisted affine Kac--Moody algebra $\fg$, let us consider the following string equation:
\begin{align}
\left(\sum_{i\in \mathscr{E}_{>0} }\left( \frac{i+ h}{ h} t_{i+
h}-\dt_{i,1}\right)\frac{\pd }{\pd t_{i}}+\frac{1}{2 h}\sum_{i,j\in
\mathscr{E}_{>0};~ i+j= h}i j t_i t_j\right)\tau(\bt) =0. \label{tauste}
\end{align}
%where $h=\sum_{i=0}^{\ell}k_i$ is the Coxeter number for $\fg$.
A formal series tau function $\tau(\bt)$ that solves the string equation is called
\emph{topological}, and we denote it as $\tau^{\mathrm{top}}(\bt)$.

The string equation gives us some useful information on the initial
value of the topological tau function $\tau^{\mathrm{top}}(\bt)$.
\begin{lem}
    The topological tau function $\tau^{\mathrm{top}}(\bt)$ satisfies
    \begin{align}
&\left.\frac{\p^{k+1}\log\tau^{\mathrm{top}}}{\p t_{j}\p
{t_1}^{k}}\right|_{\bt=0}=\dt_{j, h-1}\dt_{k,2}\frac{h-1}{h}, \quad
j\in\mathring{\mathscr{E}}, ~ k\ge 0. \label{taut03}
\end{align}
\end{lem}
\begin{prf}
It is actually easy to see from the string equation \eqref{tauste} that
\begin{align}
&\left.\frac{\p\log\tau^{\mathrm{top}}}{\p {t_1}}\right|_{\bt=0}=0, \label{taut01} \\
&\left.\frac{\p^{k+1}\log\tau^{\mathrm{top}}}{\p t_{j}\p
{t_1}^{k}}\right|_{\bt=0}=\dt_{j, h-1}\dt_{k,2}\frac{h-1}{h}, \quad
j\in\mathring{\mathscr{E}}, ~ k\ge 1; \label{taut02}
\end{align}
so that we are just left with proving
$$\left.\frac{\p\log\tau^{\mathrm{top}}}{\p {t_j}}\right|_{\bt=0}=0, \quad j\in\mathring{\mathscr{E}}.$$
Let us call temporary $A_1(\bt)$ and $A_2(\bt)$ the linear and
quadratic term of $\log\tau^{\mathrm{top}}$. More specifically,
because of \eqref{taut01}, we can write $A_1(\bt) = \sum_{j \in
\mathscr{E}_{>1}}a_jt_j$. Our goal is to prove $a_j = 0$ for $j\in\mathring{\mathscr{E}}$.
Using the string equation and integrating once with respect to
$t_1$, it is easy to see
$$A_2(\bt) = \sum_{j \in \mathscr{E}_{>1}}\frac{j + h}{h}a_j t_{j + h} t_1 + (\text{quadratic
terms independent of $t_1$}),$$
hence
\begin{equation}\label{ajtau}
  a_j=\frac{h}{j+h}\left.\frac{\pd^2\log\tau^{\mathrm{top}}}{\pd t_1\pd t_{j+h}}\right|_{\bt=0}, \quad j\in\mathring{\mathscr{E}}.
\end{equation}
It is shown in \cite{WuVir} that $\pd_{t_1}\pd_{t_{j+h}}\log\tau$ are up to a constant factor with the Hamiltonian densities of the Drinfeld--Sokolov hierarchy, and they are differential polynomials in $\pd_{t_1}\pd_{t_{i}}\log\tau\ (i\in\mathring{\mathscr{E}})$ with respect to $t_1$. Such differential polynomials are homogeneous if one assigns degree $j$ to $\frac{\pd}{\pd t_j}$, hence they (of degree $h+j+1$) do not contain monomials of the form $\pd_{t_1}^2\pd_{t_{h-1}}\log\tau$ (of degree $h+1$) or its powers whenever $j\in\mathring{\mathscr{E}}$.
Therefore by using \eqref{taut02} and \eqref{ajtau} we derive $a_j=0$ for $j\in\mathring{\mathscr{E}}$, which completes the proof.

\end{prf}

For the purpose of solving the string equation \eqref{tauste}, let
us recall that this can be equivalently written as equation (4.55)
of \cite{WuVir}, i.e.,
\begin{align}
\bigg(\Ta\bigg(d'_{-1}+\sum_{i\in \mathscr{E}_{>0}}\left(\frac{i  \,t_{i}}{h}
- \delta_{i,h+1}\right)\Ld_{i-h}\bigg)\Ta^{-1}
 -d_{-1}\bigg)_{<0\,[\rs^0]}=0. \label{steTa}
\end{align}
Here (we adapt the signs used in \cite{WuVir})
\begin{align}
d_{-1}=\frac{\od}{\od z},\quad d'_{-1}=\frac{\od}{\od
z}+\frac{1}{h\,z}\rho, \label{dm1}
\end{align}
with
\begin{equation}\label{rho}
\rho :=\sum_{i = 1}^\ell c_i H_i, \quad (c_1,\ldots,c_\ell) :=
(1,\ldots,1) \mathring{A}^{-1}.
\end{equation}
Recall that $\mathring{A}=(a_{i j})_{1\le i, j\le \ell}$ is the
Cartan matrix for the simple Lie algebra $\mathring{\fg}$. We remark
that $d_{-1}$ and $d'_{-1}$ are elements of certain Virasoro
algebras acting on $\fg$ (see, for instance, \S\,4.2 of
\cite{WuVir}). Clearly,
\begin{equation}\label{}
d_{-1}'-d_{-1}=\frac{1}{ h z }\rho\in \fg_{- h\,[\rs^1]},
\end{equation}
and, for any $X_k\in\fg_{k\,[\rs^1]}$,
\begin{equation}\label{dX}
[d_{-1}', X_k]=\frac{k}{ h z}X_k\in\fg_{k- h\,[\rs^1]}.
\end{equation}

\begin{lem}
Given the topological solution $\tau^{\mathrm{top}}$ of a
Drinfeld--Sokolov hierarchy, the associated functions $q$ and
$\omega$ satisfy the initial condition
\begin{equation}\label{}
\left.q\right|_{\bt=0}=0, \quad
\left.\om\right|_{\bt=0}=0.
\end{equation}
\end{lem}
\begin{prf}
First of all, we recall from \cite{WuVir} that $\omega =\mathrm{const}\cdot\partial_{t_1} \log\tau^{\mathrm{top}}$, so the initial condition for $\om$ is verified by using the previous lemma. On
the other hand, it is known in general that the matrix valued
function $q$ in \eqref{sL} is a differential polynomial in the
variables $\p_{t_j}\log\tau^{\mathrm{top}}$ ($j\in\mathring{\mathscr{E}}$) with respect
to $t_1$, see Remark~\ref{rmk-Taom}. If we assign degree $j$ to
$\frac{\p}{\p t_j}$, we claim that (each component of) the
differential polynomial $q$ has degree not higher than $h$ and
hence, using the previous lemma again, the proof will be finished.
Let us explain why the claim holds true.

Given a matrix valued differential operator whose entries are
polynomials in the derivatives of $\log\tau^{\mathrm{top}}$ with respect to the time variables, we define its
total degree to be the sum of degrees given by the principal
gradation on $\fg$ and the gradation described above. In  this way,
it can be seen from \cite{DS,WuVir} that $\bar\sL$ in Proposition
\ref{thm-dr0} is homogeneous of total degree $1$, and so is $\sL$ in
\eqref{sL}. We complete the proof by using the fact that each
component of $q$ has principal degree not lower than $1-h$.
\end{prf}

The string equation, for $\bt = 0$, gives the following equation for the initial value\footnote{Remark the notation $\hat\gamma$, indicating that we did not (yet) rescale $\Theta$ by $\lambda$. } $\hat\gamma(z) = \Theta^{-1}|_{\bt = 0}$.

\begin{lem}
The initial value $\hat\gamma$ of the topological solution satisfies the
equation
\begin{equation}\label{strgm1}
\hat\gm^{-1}(\Ld_1 - d_{-1}')\hat\gm=\Ld_1 - d_{-1}.
\end{equation}
\end{lem}
\begin{prf}
Thanks to the lemma above, the equality (coming from \eqref{sL} and \eqref{LTa})
$$\frac{\od}{\od x} + \Lambda + q = \Theta\left(\frac{\od}{\od x} + \Lambda\right)\Theta^{-1} + \omega \cdot c$$
 with $\bt=0$ yields
\begin{equation}\label{}
\Ld=\hat\gm^{-1}\Ld\hat\gm+\hat\gm^{-1}\left.(\p_x\Ta^{-1})\right|_{\bt=0},
\end{equation}
and taking the non-negative part we get $\Ld=p_+(\hat\gm^{-1}\Ld\hat\gm)$
(recall $p_\pm$ in Section~2). Since $\Lambda$ and $\Lambda_1$ are
the same up to a constant factor, we also get
\begin{equation}\label{gmLd1}
\Ld_1=p_+(\hat\gm^{-1}\Ld_1\hat\gm).
\end{equation}
On the other hand, again for $\bt=0$, the
string equation \eqref{steTa} is reduced to
\begin{equation}\label{strgm}
p_-\left(\hat\gm^{-1}\Ld_1\hat\gm-\hat\gm^{-1}d_{-1}'\hat\gm+d_{-1}\right)=0.
\end{equation}
Combining \eqref{gmLd1} with \eqref{strgm} we complete the proof.
\end{prf}

\begin{dfn}
Equation \eqref{strgm1} is called the reduced string equation for
the Drinfeld--Sokolov hierarchy \eqref{Lt2} associated to $\fg$.
\end{dfn}

The following is the key result to select the topological solution among the tau functions.
\begin{thm}\label{thm-gm}
Given $\fg$ an arbitrary untwisted affine Kac--Moody algebra, the reduced string
equation \eqref{strgm1} has a unique solution $\hat\gm$ of the
form
\begin{equation}\label{gamma}
\hat\gm=\exp\left(-\sum_{i\ge1}Y_{-i( h+1)}\right), \quad Y_{-i(
h+1)}\in \fg_{{-i( h+1)}\,[\rs^1]},
\end{equation}
in the Lie group of $\fg_{<0[\rs^0]}$.
\end{thm}
\begin{prf}
In this proof we will use several times the following decompositions derived from \eqref{dec2}:
\begin{equation}\label{dec}
\fg_{<0\,[\rs^1]}=\cH_{<0\,[\rs^1]}\oplus\left(\fg_{<0\,[\rs^1]}\cap\im\,\ad_{\Ld_1}\right),
\quad \cH_{<0\,[\rs^1]}=\ker\,\ad_{\Ld_1}\cap\fg_{<0\,[\rs^1]}.
\end{equation}

First of all we prove that, if a solution solving the reduced string
equation \eqref{strgm1} exists and it belongs to the Lie group of
$\fg_{<0[\rs^0]}$, then it is of the form \eqref{gamma}. In fact,
one writes a solution as
\[
\hat\gm=\exp\left(-\sum_{i\ge 1}\tilde{Y}_{-i}\right), \quad
\tilde{Y}_{-i}\in \fg_{-i\,[\rs^1]}\cap\fg_{<0\,[\rs^0]},
\]
and supposes that $-k$ is the greatest number not divided by $ h+1$
such that $\tilde{Y}_{-k}\ne0$. Then one has
\begin{align}
&[\tilde{Y}_{-k}, \Ld_1]=0, \\
&[\tilde{Y}_{-k-( h+1)}, \Ld_1]+[\tilde{Y}_{-k}, -d_{-1}' ]=0.
\label{Ykd}
\end{align}
It follows from the first equation that $\tilde{Y}_{-k}\in\cH$,
which implies $[\tilde{Y}_{-k}, -d_{-1}' ]\in\cH$ (in fact, one has
$[d'_{-1},\cH]\subset\cH$ because of \eqref{dX}). By using
\eqref{dec}, equation \eqref{Ykd} yields
\[
[\tilde{Y}_{-k-( h+1)}, \Ld_1]=0, \quad [\tilde{Y}_{-k}, -d_{-1}'
]=0,
\]
but the second equation implies $\tilde{Y}_{-k}=0$ due to
\eqref{dX}; hence a contradiction. Thus if a solution exists, it is of the form \eqref{gamma}.

Now we prove that the elements $Y_{-i(h+1)}$ are uniquely determined by recursion.

Let us substitute \eqref{gamma} into \eqref{strgm1} and compare the
homogeneous terms according to the principal gradation. Recall that
in the present realization of $\fg$, the parameter $z$ ``weights''
degree $h$ under the principal gradation, while both operators
$\ad_{d_{-1}'}$ and $\ad_{d_{-1}}$ are of degree $-h$. Explicitly,
we have
\begin{align}\label{}
&\deg 1 :  ~~ \Ld_1=\Ld_1, \\
& \deg -(h+1)+1 :  ~~ [Y_{-( h+1)}, \Ld_1] = \frac{1}{ h z}\rho, \label{eq1}\\
&\deg -2(h+1)+1 : [Y_{-2( h+1)}, \Ld_1] = -[Y_{-( h+1)},-d_{-1}']
- \frac{1}{2}[Y_{-(h+1)},[Y_{-( h+1)}, \Ld_1] ], \label{eq2}\\
&\deg -3( h+1)+1 :  [Y_{-3( h+1)}, \Ld_1] =  \nn\\
&\qquad\qquad  -[Y_{-2(h+1)}, -d_{-1}']  - \frac{1}{2}([Y_{-2(
h+1)},[Y_{-(h+1)}, \Ld_1] ] - [Y_{-(
h+1)},[Y_{-2( h+1)}, \Ld_1]])\nn\\
&\qquad\qquad   -\frac{1}{6}[Y_{-(h+1)},
[Y_{-(h+1)},[Y_{-(h+1)},\Ld_1]]] -\frac{1}{2}[Y_{-(h+1)},
[Y_{-(h+1)},-d_{-1}']],
\\
 & \qquad \vdots
\nn\\
&\deg -i( h+1)+1 :  [Y_{-(i+1)( h+1)}, \Ld_1] = \nn\\
&\qquad\qquad  -[Y_{-(i-1)( h+1)}, -d_{-1}']  -
\frac{1}{2}([Y_{-(i-1)( h+1)},[Y_{-( h+1)}, \Ld_1] ]+[Y_{-(
h+1)},[Y_{-(i-1)( h+1)}, \Ld_1]]) + *
\label{eq3}\\
 & \qquad \vdots \nn
\end{align}
where ``$*$'' stands for terms depending on $Y_{-k( h+1)}$ with
$k=1,2,\dots,i-2$.

The above equations can be solved recursively by using the property
\eqref{dec} as follows:
\begin{itemize}
\item[(I)] Consider equation \eqref{eq1}. Since $-(h+1)$ is
an exponent of $\fg$ and $- h$ is not, the equation determines
$Y_{-( h+1)}$ up to addition of a multiple of $\Ld_{-( h+1)}$. Such
a freedom is fixed uniquely by \eqref{eq2} restricted to $\cH$:
\begin{equation}\label{}
\left.\left(-[Y_{-( h+1)},-d_{-1}']-\frac{1}{2}[Y_{-( h+1)},[Y_{-(
h+1)}, \Ld_1] ]\right)\right|_\cH =0
\end{equation}
(here again we have used the property $[d'_{-1},\cH]\subset\cH$).
\item[(II)] Now let us assume that $Y_{-k( h+1)}$ with $k=1,2,\dots,i-1$ are
already known. We want to show that $Y_{-i(h+1)}$ is
uniquely determined.\\
Clearly, there are four cases: \subitem(a) $-i(h+1) \not\in \mathscr{E},
-i(h+1)+1 \not\in \mathscr{E}$; \subitem(b) $-i(h+1) \not\in \mathscr{E}, -i(h+1)+1 \in
\mathscr{E}$; \subitem(c) $-i(h+1) \in \mathscr{E}, -i(h+1)+1 \not\in \mathscr{E}$;
\subitem(d) $-i(h+1) \in \mathscr{E}, -i(h+1)+1 \in \mathscr{E}$.\\
Let's start with uniqueness (supposing that $Y_{-i(h+1)}$ exists, which will be proved later). For the cases (a) and (b), it follows from the fact that the map $\ad_{\Lambda_1}$ restricted to $\fg_{-i(h+1)[\rs^{1}]}$ is invertible.\\
For the cases (c) and (d), $Y_{-i( h+1)}$ is determined up to a multiple of $\Ld_{-i( h+1)}$,
and the freedom is fixed by
\begin{equation}
\left.\left(\hbox{ r.h.s. of equation of degree $-(i+1)( h+1)+1$
}\right)\right|_\cH=0,
\end{equation}
noting $-(i+1)( h+1)+1\in \mathscr{E}$.\\
Now for the existence: for the cases (a) and (c) the equation
\begin{equation}\label{eq4}
\left.\left(\hbox{ r.h.s. of equation of degree $-i( h+1)+1$
}\right)\right|_\cH=0,
\end{equation}
is automatically satisfied, hence, without the restriction, it belongs to the image of $\ad_{\Lambda_1}$.\\
For the cases (b) and (d) the condition \eqref{eq4} holds due to the argument we used in fixing the freedom for $Y_{-(i-1)( h+1)}$.
\end{itemize}
Therefore the theorem is proved.
\end{prf}
\begin{rmk}
Observe that the reduced string equation \eqref{strgm} can be
converted to
\begin{equation}\label{strgm3}
\left(\Ld_1-\frac{\od}{\od z}-\frac{1}{
hz}\rho\right)(\hat\gm)=\hat\gm\left(\hat\gm^{-1}\Ld_1\hat\gm\right)_{+}.
\end{equation}
Given its solution $\hat\gamma$ \eqref{gamma}, consider the $\lambda$--rescaled $\gamma(z) := \exp \left( - \sum_{i \geq 1} \lambda^{i(h+1)}Y_{-i(h+1)} \right)$ and $W_{\gamma} := \gamma H_+^{(n)}$.
Since  $\gm\left(\gm^{-1}\Ld_1\gm\right)_{+}\in W_{\gm}$ (each
column), we obtain the following (matrix version) of the
Kac--Schwarz \cite{KacSchwarz} conditions:
\begin{equation}\label{ste}
z W_{\gm}\subseteq W_{\gm}, \quad \mathcal{R}_\fg W_{\gm}\subseteq W_{\gm}
\end{equation}
with
\begin{equation}\label{Rg}
\mathcal{R}_\fg := \frac{\od}{\od z} + \frac{1}{ hz}\rho - \Ld_1.
\end{equation}

In the $A_{n-1}$ case, those operators were obtained by Kac and
Schwarz in the appendix of \cite{KacSchwarz} (together with some
suggestions about the generalization to the simply-laced cases).
Indeed, using the isomorphism (recall $H^{(n)}$ in Section~2)
\begin{align}\label{iso}
\Xi: H^{(n)} ~\to& ~~ H^{(1)}, \nn\\
\left(
  \begin{array}{c}
    f_1(z) \\
    f_2(z) \\
    \vdots \\
    f_n(z) \\
  \end{array}
\right) ~\mapsto&~~ f_1({z}^n)+{z}
f_2({z}^n)+\dots+{z}^{n-1}f_n({z}^n),
\end{align}
one can verify that the image $\widetilde W_\gamma := \Xi(W_\gamma)$ satisfies the invariance relations
\begin{equation}\label{ste2}
{z}^n \widetilde W_\gamma \subseteq \widetilde W_\gamma, \quad
\widetilde{\mathcal{R}}_n\widetilde W_\gamma\subseteq \widetilde
W_\gamma
\end{equation}
with
\begin{equation}\label{}
\widetilde{\mathcal{R}}_n=\frac{1}{n\,{z}^{n-1}}\frac{\od}{\od{z}}-\frac{n-1}{2n\,{z}^n}-{z},
\end{equation}
hence recover the well-known result in \cite{KacSchwarz}.
\end{rmk}
\begin{rmk}
It is natural to ask if the above procedure to compute the topological solutions can be extended to Drinfeld--Sokolov hierarchies associated to affine Kac--Moody algebras of twisted type. The answer is negative. In fact, in the twisted case a constraint of the form \eqref{tauste} is too strong and it is not satisfied (at least in general) by any solution, so that the notion of ``topological solution'' does not make sense. We will study solutions of Drinfeld--Sokolov hierarchies satisfying additional constraints different from the string equation in a separate publication.
\end{rmk}

\section{Topological solutions of Drinfeld--Sokolov hierarchies}

We want to combine the results of the previous two sections to
investigate the topological tau functions of the Drinfeld--Sokolov
hierarchies.

\subsection{An algorithm to compute topological tau functions}

%To this aim,
%Let us consider the entries of the matrices $g(\bt;z)$ and
%$\gamma(z)$ in the section above as graded formal series, and
%introduce a counting parameter $\lambda$ as follows:
%\begin{align}\label{resc1}
%&\gm(z)=\exp\left(- \sum_{i\ge1}\,\ld^{i(h+1)} Y_{-i(h+1)}\right), \quad g(\bt;z)=\exp\left(\sum_{j\in
%\mathscr{E}_{>0}}\ld^j t_j\Ld_j\right).
%\end{align}
%That is to say, the entries of the logarithm of $g(\bt;z)$ and
%$\gamma(z)$ are given $\ld$-degree equal to the absolute value of
%their principal degree, which implies that they are $h$--admissible in the sense of Definition
%\ref{equivassumption}, with $h$ the Coxeter number of the Kac--Moody algebra we are considering.
Combining the results of the two previous sections, we arrive to the following theorem :
\begin{thm}\label{mainthm}
Let $\fg$ be an arbitrary untwisted affine Kac--Moody algebra. Its
topological solution is given by
\begin{equation}\label{toptaudet}
\log\tau^{\mathrm{top}} = \kappa\log\det (\Id -
H(J_\gamma)\widetilde H(J_\gamma^{-1})).
\end{equation}
Here $\kappa$ is as in \eqref{tautauDS}, $J_\gamma(\bt,z) =
g^{-1}(\bt;z)\gamma(z)$  and $\gamma(z)$ is the solution (graded
with $\lambda$) of the reduced string equation \eqref{strgm1}.
\end{thm}

\begin{prf}
According to Theorems \ref{thmtaufunction} and \ref{thm-tautauDS},
the right hand side of \eqref{toptaudet} is a tau function of the
Drinfeld--Sokolov hierarchy, which corresponds to the (rescaled) initial value
$\gamma(z)$. According to Theorem \ref{thm-gm}, such an initial
value is the one determined by the reduced string equation \eqref{strgm1},
therefore the theorem is proved.
\end{prf}

%Now we have an algorithm to compute the topological tau function of
%the Drinfeld--Sokolov hierarchy associated to an arbitrary untwisted
%affine Kac--Moody algebra $\fg$. The algorithm is as follows.
%\begin{description}
%\item[~~Step I] We solve the reduced string equation \eqref{strgm1} to get $\gm(z) = \sum_{i\ge0}\gm_{-i}z^{-i}$
%as in \eqref{gamma}.
%\item[~~Step II] Let $J_\gamma(\bt;z):=g(\bt;z)^{-1}\gamma(z)$ and consider the two associated Hankel matrices $H(J_\gm)$ and $\tilde{H}(J_\gm^{-1})$ as in \eqref{ToeplitzHankel}.
%\item[~~Step III] Introduce the $\N\times\N$ matrix
%\[
%Q := H(J_\gm)\tilde{H}(J_\gm^{-1}),
%\]
%and compute the topological tau function as
%\begin{equation}\label{logtaudet}
%\log\tau^{\mathrm{top}} = \kappa\log\det (\Id -
%Q)=-\kappa\sum_{i=1}^\infty\frac{1}{i} \mathrm{Tr}Q^i.
%\end{equation}
%\end{description}

This theorem suggests an algebraic algorithm to compute
$\log\tau^{\mathrm{top}}$ up to an arbitrary order in $\lambda$ as
follows.

Let us consider the following expansions:
\begin{align}\label{}
& J_\gm(\bt;z)=g(\bt;z)^{-1}\gm(z) = \sum_{i\in\Z}J_i(\bt)z^i, \quad
J_i(\bt) = \sum_{k\ge\max\{0,-i\} }g_{i+k}(-\bt)\gm_{-k}, \\
&
J_\gm^{-1}(\bt;z)=\gm(z)^{-1}g(\bt;z)=\sum_{i\in\Z}\tilde{J}_i(\bt)z^i,
\quad \tilde{J}_i(\bt) = \sum_{k\ge\max\{0,-i\}
}(\gm^{-1})_{-k}\,g_{i+k}(\bt)
\end{align}
(the integer subscripts mean to take the coefficients of the
corresponding powers in $z$). Given a positive integer $N$, let
\begin{equation}\label{QN}
R_N := \left(\begin{array}{cccc}
J_1 & J_2 & \ldots & J_{N}\\
&&&\\
J_2 & J_3 & \ldots & J_{N+1}\\
&&&\\
   \vdots       & \vdots & \ddots & \vdots\\
&&&\\
J_N & J_{N+1} & \ldots & J_{2N-1}
\end{array}\right)\left(\begin{array}{cccc}
\tilde{J}_{-1} & \tilde{J}_{-2} & \ldots & \tilde{J}_{-N}\\
&&&\\
\tilde{J}_{-2} & \tilde{J}_{-3} & \ldots & \tilde{J}_{-N-1}\\
&&&\\
   \vdots       & \vdots & \ddots & \vdots\\
&&&\\
\tilde{J}_{-N} & \tilde{J}_{-N-1} & \ldots & \tilde{J}_{-2N+1}
\end{array}\right),
\end{equation}
Up to a certain $\ld$-degree, $R_N$ is the (block) $N\times N$ upper left
minor of the infinite matrix $R:=H(J_\gm)\tilde{H}(J_\gm^{-1})$, and
its entries are the only ones needed to compute
$\log\tau^{\mathrm{top}}$ up to order $(N+1)h + 1$, as shown in the
following proposition.
\begin{prp}\label{comp}
The topological tau function $\tau^{\mathrm{top}}$ satisfies
\begin{equation}\label{}
\deg_\ld\left(\log\tau^{\mathrm{top}}-T_N\right) \ge (N+1)h+1,
\end{equation}
where $N\ge1$ and
\begin{equation}\label{TN}
T_N=-\kappa\sum_{i=1}^{i_N}\frac{1}{i}\mathrm{Tr} {R_N}^i, \quad i_N
:= \left[\frac{(N+1) h+1}{h+2} \right].
\end{equation}
\end{prp}
\begin{prf} First of all, let us compute the $\ld$-degree of the entries of the
coefficients $\gamma^{\pm 1}_{-i}, g_{i}$. For any $k \ge 1$,
obviously we have
\begin{equation}
    \deg_\ld(\gm^{\pm 1}_{-k})_{ij} \geq h+1, \quad \deg_\ld(g_k(\bt))_{ij} \geq 1.
\end{equation}
Moreover, recalling the homogeneous realization of the Kac--Moody algebra $\fg$, the
parameter $z$ corresponds to principal degree $h$, while a constant
matrix has principal degree between $-h$ and $h$ (here we extend the principal degree of the Lie algebra to the associated matrix algebra). For this reason, we have
\begin{equation}\label{gmep}
\left.\begin{array}{l}
    \deg_\ld(\gm_{-k})_{ij}\\
    \\
    \deg_\ld(\gm^{-1}_{-k})_{ij}\\
    \\
    \deg_\ld(g_{k}(\bt))_{ij}
\end{array}\right\} \ge (k-1)h, \quad \forall\; 1\leq i,j\leq n, \quad k\ge 2.
\end{equation}

For $k\ge1$, one has
\begin{align}\label{}
&\deg_\ld (J_k(\bt))_{ij}=\deg_\ld
(g_k(-\bt)\gm_0+g_{k+1}(-\bt)\gm_{-1}+\dots)_{ij} \ge \max\{(k-1)h, 1\}, \\
&\deg_\ld(\tilde{J}_{-k}(\bt))_{ij}=\deg_\ld(\,(\gm^{-1})_{-k}g_0(\bt)
+(\gm^{-1})_{-k-1}g_1(\bt)+\dots)_{ij}\ge \max\{(k-1)h, h+1\}.
\end{align}
Observing \eqref{QN}, every entry in the last $n$ rows of $R_{N+1}$
has $\ld$-degree higher than $(N+1) h$. Then by taking the Laplace
expansion of $\det(\Id-R_{N+1})$ with respect to the last $n$ rows,
we derive
\begin{align}\label{}
\deg_\ld\left(\det(\Id-R_{N+1})-\det(\Id-R_N)\right) \ge (N+1) h +1.
\end{align}
Consequently,
\begin{equation}\label{QQN}
\deg_\ld\left(\det(\Id-R)-\det(\Id-R_N)\right) \ge (N+1) h +1.
\end{equation}

On the other hand, each entry in $R_N$ has $\ld$-degree greater or
equal than $h+2$, hence we have
\begin{equation}\label{}
\deg_\ld\mathrm{Tr} ({R_N}^k) \ge k(h+2).
\end{equation}
For the purpose
\[
(k+1)( h+2) \geq (N+1) h+1,
\]
one needs
\[
k \ge \frac{(N+1) h+1}{h+2} - 1.
\]
Hence, for such a $k$,
\begin{equation}\label{trQN}
\deg_\ld\left( \log\det(\Id-R_N)+\sum_{i=1}^k\mathrm{Tr}\frac{1}{i}
{R_N}^i\right)=\deg_\ld\left( -\sum_{i\ge k+1}\mathrm{Tr}\frac{1}{i}
{R_N}^i\right) \ge (N+1) h+1.
\end{equation}

Therefore, taking \eqref{QQN} and \eqref{trQN} together, %we  deduce,
%for $i_N$ given in \eqref{TN},
%\begin{align}\label{}
%\deg_\ld\left(\log\tau^{\mathrm{top}}-T_N\right)=&\deg_\ld\left(
%\log\tau^{\mathrm{top}}-\kappa\det(\Id-Q_N))+(\kappa\det(\Id-Q_N)-T_N)\right)
%>N h+1. \label{tauTNord}
%\end{align}
the proposition is proved.
\end{prf}

\subsection{Reductions of topological tau functions}

Inspired by the results in \cite{LRZ}, let us consider the following
situation. Given a simple Lie algebra $\mathring \fg$ together with
an automorphism $\sigma$, suppose that the invariant subalgebra
$\mathring \fg^{\sigma}$ is also a simple Lie algebra. It is known
(see \cite{Kac}) that $\mathring{\fg}$ and $\sigma$ can be chosen
(up to isomorphisms) as follows:
$$
\begin{array}{rllll}
    D_{m+1} &:& \sigma(K_i) = K_i \, (1 \leq i \leq m-1), \; \sigma(K_m) = K_{m+1}, \;
    \sigma(K_{m+1}) = K_{m};\\
    A_{2m - 1} &:& \sg(K_i) = K_{2m - i}\, (1 \leq i \leq 2 m-1);\\
    E_6 &:& \sg(K_i) = K_{6 - i}\, (i = 1,2,4,5), \; \sg(K_j) = K_j\, (j = 3,6); \\
    D_4 &:& \sg(K_1) = K_3, \; \sg(K_2) = K_2, \; \sg(K_3) = K_4, \; \sg(K_4) = K_1,
\end{array}
$$
where $K_i = E_i,H_i,F_i$ (Weyl generators for $\mathring{\fg}$). In this
way, the invariant subalgebra $\mathring \fg^\sigma$ is respectively
of type $B_m,C_m,F_4,G_2$.

Note that $\sigma$ extends naturally to an automorphism on the
corresponding affine Kac--Moody algebra $\fg$. Accordingly, the
invariant subalgebra $\fg^\sg$ is the untwisted affine Kac--Moody
algebra related to $\mathring \fg^\sigma$. Moreover, it can be seen
that $\mathcal H^\sigma \subseteq \mathcal H$ is the principal
Heisenberg subalgebra of $\fg^\sg$.
\begin{thm}\label{reductthm}
    Let $\fg$ and $\fg^\sg$ be as above and denote with $\tau^{\mathrm{top}}$
    and $\tau^{\sg, \mathrm{top}}$ the corresponding topological tau functions. Then
    \begin{equation}\label{}
\tau^{\mathrm{top}}(g) =\tau^{\sg, \mathrm{top}}(g) \quad \forall\,
g \in \exp\left(\mathcal H_{>0\,[\rs^1]}^\sigma\right).
\end{equation}
\end{thm}
\begin{prf}
    Because of Theorem \ref{mainthm}, it suffices to prove that
    the reduced string equation \eqref{strgm1} on $\fg$ and $\fg^\sg$
    has the same solution. Indeed, one can verify (see \S 7.9 of \cite{Kac})
    $\sigma(\Lambda_1) = \Lambda_1$ and $\sigma(\rho) = \rho$,
    so that the reduced  string equation \eqref{strgm1} is invariant with
    respect to $\sigma$. This concludes the theorem.
\end{prf}

Note that this theorem agrees with the results obtained in
\cite{LRZ}.

\section{Examples: explicit expansions of topological
solutions}\label{sec-expan}

By using Theorem~\ref{mainthm}, we can compute the topological solutions of the Drinfeld--Sokolov
hierarchy associated to any untwisted affine Kac--Moody algebra $\fg$, based on the data listed in the
appendix, see \cite{DS, Kac}. There, generators for the principal
subalgebras are normalized as in \cite{WuVir}.
%\textcolor{blue}{
In order to obtain the generating functions \eqref{potential} for Gromov--Witten/FJRW invariants,
we still need to introduce a parameter $\ep$ to count the genus and do some normalization of the variables. Let us consider the $r$-spin case with the normalization described as in \cite{Witten4} (see equations (1.5.1)--(1.6.3) therein). More precisely, let $\sL$ in \eqref{sL} be replaced by
\[
\sL^\ep=\frac{\ep\sqrt{-1}}{\sqrt{h}}\frac{\od}{\od x}+\Ld+q
\]
and in the definition of flows $\pd/\pd t_j$ in \eqref{Tat} the generators $\Ld_j$ are replaced by
\begin{equation}\label{Ldjep} \Ld_j
\mapsto \left( \frac{\ep\sqrt{-1}}{\sqrt{h}}\right)^{-1}\Ld_j, \quad j\in \mathscr{E}_{>0}.
\end{equation}
Rescaling also by $\lambda$, the matrices $\gamma(z)$ solving the reduced string equation and $g(\bt;z)\in G^a_+$ in \eqref{Gaplus} read
\begin{align}\label{gmgdeg}
&\gm(z)=\exp\left(-\sum_{i\ge1}\left(\frac{\ep\sqrt{-1}}{\sqrt{h}}\right)^i\ld^{i(h+1)} Y_{-i(h+1)}\right),
\quad g(\bt;z)=\exp\left(\sum_{j\in
\mathscr{E}_{>0}} \ld^j t_j \left( \frac{\ep\sqrt{-1}}{\sqrt{h}}\right)^{-1}\Ld_j\right),
\end{align}
and they are $h$-admissible.
Moreover, we let
  \begin{equation}\label{t2q}
%\lambda=1, \quad
t_{ h k+m_\al}=(-1)^{k+1}\frac{\Gm\left(\frac{m_\al}{h}\right)}{ h\,
\Gm\left(k+1+\frac{m_\al}{h}\right)}q_{\al,k}\quad \hbox{ with }
\al=1, 2, \dots, \ell; ~ k\ge0;
\end{equation}
here we recall $\mathring{\mathscr{E}}=\{m_1, m_2, \dots, m_\ell\}$.  Now the string equation \eqref{tauste} (setting temporarily $\lambda = 1$)  becomes
\begin{equation}\label{}
\sum_{\al=1}^\ell\sum_{k=0}^\infty\left(q_{\al,k+1}-\dt_{\al,1}\dt_{k,0}\right)
\frac{\pd\log\tau}{\pd q_{\al,k}}+
\frac{1}{2\ep^2}\sum_{\al=1}^{\ell}q_{\al,0}q_{\ell-\al,0}=0.
\end{equation}
Note that the $\ep$ factor in the equation appears because of \eqref{Ldjep} and the construction of Virasoro symmetries in \cite{WuVir}.
We remark that, especially for the cases that are not of $A_\ell$ type, this kind of normalization is not the only one present in the literature. This is why the coefficients reported below are equal up to a sign to the ones in \cite{LRZ} for $D_4$, and a different normalization had been chosen in \cite{HKW} for $E_6$ so to fit with \cite{FJR} .
%}

Below are explicit
results for some examples.

\subsection{The $A_1$ case}
The reduced string equation \eqref{strgm1} with the parameters
$\lambda$ and $\ep$  reads (we write $\tilde{\ep}={\ep\sqrt{-1}}/{\sqrt{2}}$ to simplify the notations)
\begin{equation}\label{}
\gm(z)^{-1}\left(\frac{\ld}{\tilde{\ep}} \left(
        \begin{array}{cc}
          0 & z \\
          1 & 0 \\
        \end{array}
      \right)-\frac{1}{\ld^2}\frac{\od}{\od z}+\frac{1}{2\ld^2 z}\left(
        \begin{array}{cc}
          -1/2 & 0 \\
          0 & 1/2 \\
        \end{array}
      \right)\right)\gm(z)=\frac{\ld}{\tilde{\ep}} \left(
        \begin{array}{cc}
          0 & z \\
          1 & 0 \\
        \end{array}
      \right)-\frac{1}{\ld^2}\frac{\od}{\od z}.
\end{equation}
This equation is solved by
\begin{align}\label{}
\gm(z)=&\Id+\left(
\begin{array}{ll}
 0 & \frac{7}{48} \\
 0 & 0
\end{array}
\right)\tilde{\ep}\,\ld^3\,z^{-1}+\left(
\begin{array}{ll}
 0 & 0 \\
 -\frac{5}{48} & 0
\end{array}
\right)\tilde{\ep}\,\ld^3\,z^{-2}+\left(
\begin{array}{ll}
 \frac{385}{4608} & 0 \\
 0 & -\frac{455}{4608}
\end{array}
\right)\tilde{\ep}^2\ld^6 z^{-3} \nn\\
&+\left(
\begin{array}{ll}
 0 & \frac{95095}{663552} \\
 0 & 0
\end{array}
\right)\tilde{\ep}^3\ld^9 z^{-4}+\left(
\begin{array}{ll}
 0 & 0 \\
 -\frac{85085}{663552} & 0
\end{array}
\right)\tilde{\ep}^3\ld^9 z^{-5}\nn\\
&+\left(
\begin{array}{ll}
 \frac{37182145}{127401984} & 0 \\
 0 & -\frac{40415375}{127401984}
\end{array}
\right)\tilde{\ep}^4\ld^{12} z^{-6}+\left(
\begin{array}{ll}
 0 & \frac{5763232475}{6115295232} \\
 0 & 0
\end{array}
\right)\tilde{\ep}^5\ld^{15} z^{-7}+\dots
\end{align}
Indeed, one can also give a closed formula for the Fourier coefficients $\gamma_{-i}$ reading
\begin{align}\label{}
\gm_{-i}%=\left(
%        \begin{array}{cc}
%          a_{2i} & b_{2i} \\
%          \\
%          a_{2i-1} & b_{2i-1} \\
%        \end{array}
%      \right)
=\left(
    \begin{array}{cc}
      a_{2i} & -\frac{4i+3}{4i+1}a_{2i+1} \\
      \\
      a_{2i-1} & -\frac{4i+1}{4i-1}a_{2i}\\
    \end{array}
  \right)
\end{align}
where, for $k\ge0$,
\begin{equation}\label{}
a_{3k+1}=a_{3k+2}=0, \quad a_{3k}=\tilde{\ep}^k\ld^{3k}\left(-\frac{3}{4}\right)^k
\frac{\Gm(k+\frac{5}{6})\,\Gm(k+\frac{1}{6})}{k!\,\Gm(\frac{5}{6})\,\Gm(\frac{1}{6})}.
\end{equation}
These coefficients appear in the large-value asymptotic expansion of
the Airy function (see, e.g., \cite{KacSchwarz}):
\[
Ai(z)\sim \frac{e^{-\frac{2}{3}z^{\frac{3}{2}}}}{4\pi^{\frac{3}{2}}
z^{\frac{1}{4}}}\sum_{k=0}^{\infty}\left(-\frac{3}{4}\right)^k
\frac{\Gm(k+\frac{5}{6})\,\Gm(k+\frac{1}{6})}{k! \, z^{\frac{3}{2}k}
}, \quad |\arg(z)|<\pi.
\]

Using Proposition \ref{comp} we get
\begin{small}
\begin{align*}
\log\tau^{\mathrm{top}}=&\frac {1}{{\tilde{\ep}}^{2}}
\bigg(\frac{1}{12}{t_{{1}}}^{3}{\ld}^{6}+\frac{1}{8}{t
_{{1}}}^{3}t_{{3}}{\ld}^{12}+ \left( {\frac
{5\,{t_{{1}}}^{4}t_{{5}}}{64}}+\frac{3}{16}{t_
{{1}}}^{3}{t_{{3}}}^{2} \right) {\ld}^{18} + \left( { \frac
{45\,{t_{{1}}}^{4}t_{{3}}t_{{5}}}{128}}+{\frac {7\,{t_{{1}}}^{5}
t_{{7}}}{128}}+{\frac {9\,{t_{{1}}}^{3}{t_{{3}}}^{3}}{32}} \right) {
\ld}^{24} \bigg) \nn\\
&+ \frac{1}{16}t_{{3}}{\ld}^{6}+ \left( {\frac {3\,{t_{{
3}}}^{2}}{64}}+{\frac {5\,t_{{1}}t_{{5}}}{32}} \right) {\ld}^{12}+
\left( {\frac {3\,{t_{{3}} }^{3}}{64}}+{\frac
{35\,{t_{{1}}}^{2}t_{{7}}}{128}}+{\frac {15\,t_{{1}
}t_{{3}}t_{{5}}}{32}} \right) {\ld}^{18} \nn\\
&+ \left( {\frac {105\,{t_{{1}}}^{3}t_{{9}}}{256}}+{\frac {75\,{t_{
{1}}}^{2}{t_{{5}}}^{2}}{128}}+{\frac
{315\,{t_{{1}}}^{2}t_{{3}}t_{{7}} }{256}}+{\frac
{135\,{t_{{3}}}^{2}t_{{1}}t_{{5}}}{128}}+{\frac {27\,{t
_{{3}}}^{4}}{512}} \right) {\ld}^{24} \nn\\
&+ {\tilde{\ep}} ^{2}\left({\frac {105\,{\ld}^{18}t_{{9}}}{1024}}+ \left(
{\frac {1155\,t_{{1}}t_{{11}}}{2048}}+{\frac
{1015\,t_{{5}}t_{{7}}}{2048}}+{\frac {945\,t_{{3}}t_{{9}}}{2048}}
\right) {\ld}^{24} \right)+\mathcal{O}(\ld^{30}).
\end{align*}
\end{small}
After the replacement \eqref{t2q}, and setting $\lambda = 1$, we obtain the well known expansion of the Witten-Kontsevich tau function
\begin{equation}\label{}
\log\tau^{\mathrm{top}}=\sum_{g \geq 0}
\epsilon^{2g-2}\mathcal{F}_g,
\end{equation}
with (denote $q_k=q_{1,k}$)
\begin{small}
\begin{align*}
\mathcal{F}_0=&\frac{{q_{{0}}}^{3}}{6}
+\frac{{q_{{0}}}^{3}q_{{1}}}{6} + \bigg(
\frac{1}{24}{q_{{0}}}^{4}q_{{2}}+\frac{1}{6}{q_{{0}}}^{3}{q_{{1}}}^{2}
 \bigg)
  +\bigg(
\frac{1}{8}{q_{{0}}}^{4}q_{{1}}q_{{2}}+\frac{1}{120}{q_{{0}}}^{5}q_{{3}}
+\frac{1}{6}{q_{{0}}}^{3}{q_{{1}}}^{3} \bigg)   \\
&+  \bigg(
\frac{1}{30}q_{{3}}{q_{{0}}}^{5}q_{{1}}+\frac{1}{4}{q_{{0}}}^{4}{q_{{1}}}^
{2}q_{{2}}+\frac{1}{40}{q_{{0}}}^{5}{q_{{2}}}^{2}+\frac{1}{6}{q_{{0}}}^{3}{q_{{1}}
}^{4}+{\frac {1}{720}{q_{{0}}}^{6}q_{{4}}} \bigg)  +
\mathrm{h.d.t.},
\\
\mathcal{F}_1=& \frac{q_{{1}}}{24} + \bigg(
\frac{1}{48}{q_{{1}}}^{2}+\frac{1}{24}q_{{0}}q_{{2}}
\bigg) + \bigg( {
\frac{1}{72}}{q_{{1}}}^{3}+\frac{1}{48}{q_{{0}}}^{2}q_{{3}}+\frac{1}{12}q_{{0}}q_{
{1}}q_{{2}} \bigg)  \\
&
 +\bigg(
{\frac{1}{144}}{q_{{0}}}^{3}q_{{4}}+\frac{1}{24}{q_{{0}}}^{2}{q_{{2}
}}^{2}+\frac{1}{16}{q_{{0}}}^{2}q_{{1}}q_{{3}}+\frac{1}{8}{q_{{1}}}^{2}q_{{0}}q_{{
2}}+{\frac {1}{96}{q_{{1}}}^{4}} \bigg) +\bigg( {\frac
{1}{120}{q_{{1}}}^{5}} +{\frac {1}{576}{q_{{0}}}^{4}q_{{5}}}
\\
&+\frac{1}{8}q_{{3}}{q_{{1}}}^{2}{q_{{0}}}^{2}
+\frac{1}{36}{q_{{0}}}^{3}q_{{4}}q_
{{1}}+{\frac{7}{144}}q_{{3}}{q_{{0}}}^{3}q_{{2}}+\frac{1}{6}{q_{{0}}}^{2}
{q_{{2}}}^{2}q_{{1}}+\frac{1}{6}q_{{0}}{q_{{1}}}^{3}q_{{2}}
\bigg)  + \mathrm{h.d.t.},
\\
\mathcal{F}_2=&{\frac {q_{{4}}}{1152}} + \bigg( {\frac
{1}{1152}}q_{{0}}q_{{5}}+{\frac {29}{ 5760}}q_{{2}}q_{{3}}+{\frac
{1}{384}}q_{{1}}q_{{4}} \bigg) + \bigg( {\frac
{29}{5760}}q_{{0}}{q_{{3}}}^{2}+{\frac {7}{1440}}{q_{{2}}}^{3} \\
& +{\frac {11}{1440}}q_{{2}}q_{{0}}q_{{4}}+{\frac
{1}{288}}q_{{0}}q_{{1}}q_{{5}}+{\frac
{29}{1440}}q_{{3}}q_{{1}}q_{{2}}+{\frac
{1}{2304}}{q_{{0}}}^{2}q_{{6}}+{\frac {1}{192}}q_{{4}}{q_{{1}}}^{2}
 \bigg) + \mathrm{h.d.t.},
 \\
 \mathcal{F}_3=&\frac {1}{82944}q_{{7}} + \mathrm{h.d.t.}
\end{align*}
\end{small}
Here and below, the notation `h.d.t.' stands for higher-degree terms with respect to
 $\deg
q_{\al,k}=\deg
t_{h k+m_\al}=h k+m_\al$.
% One can analyzes the principal degrees of the exponents in
%\eqref{gmgdeg}, then sees that $\ld^{l}$ collects homogeneous
%polynomials of degree $l$ under the grading $\deg
%q_k=\dfrac{2k+1}{3}$.

\subsection{Other examples}
In the following examples, we just report the matrices $\Lambda_1$
and $\rho$, but omit the solution $\gamma(z)$ of the reduced string
equation \eqref{strgm1}. We also omit the superscript $(W,G)$ in \eqref{potential}  % The logarithm of the topological tau
%function
%\begin{equation}\label{}
%\log\tau^{\mathrm{top}}=\sum_{g \geq 0} \epsilon^{2g-2}\mathcal{F}_g
%\end{equation}
% will be written under the following rescaling:
%  \begin{equation}\label{}
%\ep\mapsto \frac{\ep}{\sqrt{h}},\quad \lambda^{2(h+1)} \mapsto
%\lambda,
%\end{equation}
%\begin{equation} \label{t2q}
%t_{ h k+m_\al}\mapsto \frac{\Gm\left(\frac{m_\al}{h}\right)}{ h
%\Gm\left(k+1+\frac{m_\al}{h}\right)}q_{\al,k}\quad \hbox{ with }
%\al=1, 2, \dots, \ell,
%\end{equation}
% \begin{equation}\label{}
%\ep\mapsto (-1)^{\frac{2}{h}-\frac{1}{2}}\frac{\ep}{\sqrt{h}},\quad
%\lambda^{2(h+1)} \mapsto (-1)^{\frac{2}{h}} \lambda,
%\end{equation}
%\begin{equation} \label{t2q}
%t_{ h k+m_\al}\mapsto (-1)^{\frac{m_\al +1}{h}}
%\frac{\Gm\left(\frac{m_\al}{h}\right)}{ h
%\Gm\left(k+1+\frac{m_\al}{h}\right)}q_{\al,k}\quad \hbox{ with }
%\al=1, 2, \dots, \ell,
%\end{equation}
%where $m_1, m_2, \dots, m_\ell$ are the exponents between $0$ and $
%h$.

\begin{itemize}
    \item {\bf The $A_2$ case}
            \begin{equation}\label{}
                \Ld_1=\left(
                            \begin{array}{ccc}
                            0 & 0 & z \\
                            1 & 0 &0\\
                            0 & 1 & 0
                            \end{array}
                            \right), \quad \rho=\left(
                            \begin{array}{ccc}
                            -1 & 0 & 0\\
                            0 & 0 & 0 \\
                            0 & 0 & 1 \\
                            \end{array}
                        \right).
                \end{equation}
\begin{small}
\begin{align*}
    \mathcal{F}_0 =& \frac{1}{2}\,{q_{{1,0}}}^{2}q_{{2,0}} + \bigg( \frac{1}{2}\,q_{{1,1}}q_
{{2,0}}{q_{{1,0}}}^{2}+{\frac
{{q_{{2,0}}}^{4}}{72}}+\frac{1}{6}\,q_{{2,1}}{q_ {{1,0}}}^{3} \bigg)
 + \bigg(\frac{1}{18}\,{q_{{2,0}}}^{3}q_{{1
,0}}q_{{2,1}}+\frac{1}{6}\,q_{{2,0}}q_{{1,2}}{q_{{1,0}}}^{3}
\\
&+\frac{1}{2}\,{q_{{1,1}}}^
{2}{q_{{1,0}}}^{2}q_{{2,0}}+\frac{1}{3}\,q_{{2,1}}q_{{1,1}}{q_{{1,0}}}^{3}+\frac{1}
{36}\,{q_{{2,0}}}^{4}q_{{1,1}}+\frac{1}{24}\,{q_{{1,0}}}^{4}q_{{2,2}}
\bigg)  + \bigg(
\frac{1}{2}\,{q_{{1,0}}}^{3}q_{{2,1}}{q_{{1,1}}}^{2} \\
&+\frac{1}{2}
\,{q_{{1,0}}}^{2}{q_{{1,1}}}^{3}q_{{2,0}}+\frac{1}{36}\,{q_{{2,0}}}^{4}q_{{1,2
}}q_{{1,0}}+\frac{1}{12}\,{q_{{2,0}}}^{2}{q_{{2,1}}}^{2}{q_{{1,0}}}^{2}+\frac{1}{8}\,{
q_{{1,0}}}^{4}q_{{2,2}}q_{{1,1}}+\frac{1}{8}\,{q_{{1,0}}}^{4}q_{{1,2}}q_{{2,1}
}
\\
&+\frac{1}{24}\,q_{{2,0}}q_{{1,3}}{q_{{1,0}}}^{4}+\frac{1}{36}\,{q_{{2,0}}}^{3}q_{{2,2
}}{q_{{1,0}}}^{2}+\frac{1}{6}\,{q_{{2,0}}}^{3}q_{{2,1}}q_{{1,0}}q_{{1,1}}+\frac{1}{2}
\,q_{{2,0}}q_{{1,2}}q_{{1,1}}{q_{{1,0}}}^{3}+{\frac
{{q_{{1,0}}}^{5}q_
{{2,3}}}{120}}\\
&+\frac{1}{24}\,{q_{{2,0}}}^{4}{q_{{1,1}}}^{2} \bigg)
+  \bigg(
\frac{1}{2}\,{q_{{1,0}}}^{2}{q_{{1,1}}}^{4}q_{{2,0}}+\frac{2}{3}\,{q_{{1,0
}}}^{3}q_{{2,1}}{q_{{1,1}}}^{3}+\frac{1}{30}\,{q_{{1,0}}}^{5}q_{{1,3}}q_{{2,1}
}+\frac{1}{20}\,{q_{{1,0}}}^{5}q_{{1,2}}q_{{2,2}}\\
&+\frac{1}{18}\,q_{{2,0}}{q_{{2,1}}}^{
3}{q_{{1,0}}}^{3}+\frac{1}{8}\,q_{{2,0}}{q_{{1,2}}}^{2}{q_{{1,0}}}^{4}
+{\frac
{{q_{{2,0}}}^{4}q_{{1,3}}{q_{{1,0}}}^{2}}{72}}+\frac{1}{4}\,{q_{{1,0}}}^{4}q_{
{2,2}}{q_{{1,1}}}^{2}+{\frac
{{q_{{2,0}}}^{3}q_{{2,3}}{q_{{1,0}}}^{3} }{108}}\\
&+{\frac
{q_{{2,0}}q_{{1,4}}{q_{{1,0}}}^{5}}{120}}+\frac{1}{30}\,{q_{{1,0
}}}^{5}q_{{2,3}}q_{{1,1}}+\frac{1}{3}\,{q_{{2,0}}}^{3}q_{{1,0}}q_{{2,1}}{q_{{1
,1}}}^{2}+\frac{1}{9}\,{q_{{2,0}}}^{4}q_{{1,0}}q_{{1,2}}q_{{1,1}}+\frac{1}{9}\,{q_{{2,0
}}}^{3}q_{{2,2}}q_{{1,1}}{q_{{1,0}}}^{2}\\
&+{\frac {5\,{q_{{2,0}}}^{3}q_{ {1,2}}q_{{2,1}}{q_{{1,0}}}^{2}}{36}}
+\frac{1}{12}\,{q_{{2,0}}}^{2}q_{{2,2}}q_{
{2,1}}{q_{{1,0}}}^{3}+\frac{1}{6}\,q_{{2,0}}q_{{1,3}}q_{{1,1}}{q_{{1,0}}}^{4}+
\frac{1}{3}\,{q_{{2,0}}}^{2}q_{{1,1}}{q_{{1,0}}}^{2}{q_{{2,1}}}^{2}\\
&+q_{{2,0}}q _{{1,2}}{q_{{1,1}}}^{2}{q_{{1,0}}}^{3}
+\frac{1}{2}\,{q_{{1,0}}}^{4}q_{{1,2}}q_ {{1,1}}q_{{2,1}}+{\frac
{{q_{{1,0}}}^{6}q_{{2,4}}}{720}}+{\frac {{q_{{
2,0}}}^{5}{q_{{2,1}}}^{2}}{360}}+\frac{1}{18}\,{q_{{2,0}}}^{4}{q_{{1,1}}}^{3}+
{\frac {{q_{{2,0}}}^{6}q_{{2,2}}}{1620}} \bigg)  \\
& +\mathrm{h.d.t.},
\\
    \mathcal F_1 =&
\frac{1}{12}\,q_{{1,1}} + \bigg(
\frac{1}{24}\,{q_{{1,1}}}^{2}+\frac{1}{12}\,q_{{1,2}}q_ {{1,0}}
\bigg) + \bigg(
\frac{1}{6}\,q_{{1,0}}q_{{1,2}}q_{{1,1}}
+\frac{1}{36}\,{q_{{1,1}}}^{3}+{\frac
{{q_{{2,0}}}^{2}q_{{2,2}}}{72}}+{\frac {
q_{{2,0}}{q_{{2,1}}}^{2}}{72}}\\
&+\frac{1}{24}\,{q_{{1,0}}}^{2}q_{{1,3}} \bigg) +
\bigg(
\frac{1}{4}\,q_{{1,2}}{q_{{1,1}}}^{2}q_{{1,0}}+\frac{1}{8}\,{q_{
{1,0}}}^{2}q_{{1,3}}q_{{1,1}}+\frac{1}{24}\,q_{{2,0}}q_{{1,1}}{q_{{2,1}}}^{2}+
\frac{1}{24}\,{q_{{2,0}}}^{2}q_{{2,2}}q_{{1,1}}\\
&+\frac{1}{24}\,{q_{{2,0}}}^{2}q_{{1,2}} q_{{2,1}}+{\frac
{{q_{{2,0}}}^{2}q_{{2,3}}q_{{1,0}}}{72}}+\frac{1}{18}\,q_{{2,
1}}q_{{2,0}}q_{{2,2}}q_{{1,0}}+\frac{1}{48}\,{q_{{1,1}}}^{4}+{\frac
{{q_{{2,0} }}^{3}q_{{1,3}}}{108}}+{\frac
{{q_{{2,1}}}^{3}q_{{1,0}}}{72}}\\
&+\frac{1}{12}\,{q _{{1,0}}}^{2}{q_{{1,2}}}^{2}+{\frac
{q_{{1,4}}{q_{{1,0}}}^{3}}{72}}
 \bigg) + \bigg(\frac{1}{36}\,{q_{{2,2}}}^{2}q_{{2,0}}{q_{{1,0
}}}^{2}+{\frac
{7\,q_{{1,3}}q_{{1,2}}{q_{{1,0}}}^{3}}{72}}+\frac{1}{18}\,q_{{1
,4}}{q_{{1,0}}}^{3}q_{{1,1}}\\
&+\frac{1}{18}\,{q_{{2,1}}}^{3}q_{{1,1}}q_{{1,0}}+\frac{1}{3}\,
q_{{1,2}}{q_{{1,1}}}^{3}q_{{1,0}}+\frac{1}{12}\,q_{{2,0}}{q_{{1,1}}}^{2}{q
_{{2,1}}}^{2}+\frac{1}{12}\,{q_{{2,0}}}^{2}q_{{2,2}}{q_{{1,1}}}^{2}+\frac{1}{27}\,{q_{
{2,0}}}^{3}q_{{1,3}}q_{{1,1}}\\
&+{\frac {{q_{{2,0}}}^{2}q_{{2,4}}{q_{{1,0 }}}^{2}}{144}}+{\frac
{{q_{{2,0}}}^{3}q_{{1,0}}q_{{1,4}}}{108}}+\frac{1}{3}\,{
q_{{1,0}}}^{2}{q_{{1,2}}}^{2}q_{{1,1}}+{\frac
{7\,{q_{{1,0}}}^{2}q_{{2
,2}}{q_{{2,1}}}^{2}}{144}}+\frac{1}{4}\,{q_{{1,0}}}^{2}q_{{1,3}}{q_{{1,1}}}^{2
}\\
&+\frac{1}{18}\,{q_{{2,0}}}^{2}q_{{1,0}}q_{{2,3}}q_{{1,1}}+{\frac
{5\,{q_{{2,0
}}}^{2}q_{{1,0}}q_{{1,3}}q_{{2,1}}}{72}}+\frac{1}{12}\,{q_{{2,0}}}^{2}q_{{1,0}
}q_{{1,2}}q_{{2,2}}+\frac{1}{6}\,{q_{{2,0}}}^{2}q_{{1,2}}q_{{1,1}}q_{{2,1}}\\
&+\frac{1}{24}\,
q_{{2,3}}q_{{2,1}}{q_{{1,0}}}^{2}q_{{2,0}}+\frac{1}{8}\,q_{{1,2}}{q_{{2,1}
}}^{2}q_{{2,0}}q_{{1,0}}+\frac{2}{9}\,q_{{2,2}}q_{{1,1}}q_{{1,0}}q_{{2,1}}q_{{
2,0}}+{\frac {{q_{{1,1}}}^{5}}{60}}+{\frac
{{q_{{1,0}}}^{4}q_{{1,5}}}{ 288}}\\
&+{\frac {5\,{q_{{2,0}}}^{3}{q_{{1,2}}}^{2}}{216}} \bigg) +\mathrm{h.d.t.},
\\
    \mathcal F_2 =&
\bigg({\frac {17\,{q_{{2,2}}}^{2}}{8640}}+{\frac
{q_{{2,0}}q_{{2,4} }}{864}}+{\frac {11\,q_{{2,3}}q_{{2,1}}}{4320}}
\bigg) +
 \bigg({\frac {q_{{2,0}}q_{{2,3}}q_{{1,2}}}{108}}+{\frac {19\,q_{{1,
2}}q_{{2,2}}q_{{2,1}}}{1080}}+{\frac
{q_{{2,4}}q_{{1,0}}q_{{2,1}}}{270 }}\\
 &+{\frac
{11\,q_{{2,3}}q_{{2,1}}q_{{1,1}}}{1080}}+{\frac {23\,q_{{2,0
}}q_{{1,3}}q_{{2,2}}}{2160}}+{\frac
{7\,q_{{2,3}}q_{{2,2}}q_{{1,0}}}{ 1080}}+{\frac
{q_{{2,5}}q_{{1,0}}q_{{2,0}}}{864}}+{\frac {q_{{2,0}}q_{
{2,4}}q_{{1,1}}}{216}}+{\frac
{13\,q_{{2,0}}q_{{1,4}}q_{{2,1}}}{2160}} \\
 &+{\frac
{17\,{q_{{2,2}}}^{2}q_{{1,1}}}{2160}}+{\frac {29\,q_{{1,3}}{q_
{{2,1}}}^{2}}{4320}}+{\frac {{q_{{2,0}}}^{2}q_{{1,5}}}{864}} \bigg)
  +\mathrm{h.d.t.},
\\
\mathcal F_3 =& {\frac {q_{{2,6}} }{31104}} + \mathrm{h.d.t.}
\end{align*}
\end{small}

The result coincides with that given by Zhou in \cite{Zh2}, which is
based on the validity of conjectured commutativity between some
$W$-constraint operators.

\item {\bf The $A_3$ case}
\begin{equation}\label{LdrhoA3}
\Ld_1=\left(
        \begin{array}{cccc}
          0 & 0 & 0 & z \\
          1 & 0 &0 & 0\\
          0 & 1 & 0 & 0\\
          0 & 0 & 1 & 0\\
        \end{array}
      \right), \quad \rho=\left(
        \begin{array}{cccc}
          -\frac{3}{2} & 0 & 0 & 0\\
          0 & -\frac{1}{2} & 0 & 0 \\
          0 & 0 & \frac{1}{2} & 0 \\
          0 & 0 & 0 & \frac{3}{2} \\
        \end{array}
      \right).
\end{equation}
\begin{small}
\begin{align*}
\mathcal F_0 =& \bigg(
\frac{1}{2}\,{q_{{1,0}}}^{2}q_{{3,0}}+\frac{1}{2}\,q_{{1,0}}{q_{{2,0}}}^{2}
 \bigg) + \bigg( \frac{1}{6}\,{q_{{1,0}}}^{3}q_{{3,1}}+\frac{1}{16}\,{q_{{2,0}
}}^{2}{q_{{3,0}}}^{2}+\frac{1}{2}\,q_{{1,1}}{q_{{1,0}}}^{2}q_{{3,0}}+\frac{1}{2}\,q_{{
1,1}}q_{{1,0}}{q_{{2,0}}}^{2}\\
&+\frac{1}{2}\,{q_{{1,0}}}^{2}q_{{2,0}}q_{{2,1}}
 \bigg) + \bigg( \frac{1}{6}\,{q_{{1,0}}}^{3}{q_{{2,1}}}^{2}+{
\frac
{{q_{{2,0}}}^{4}q_{{3,1}}}{96}}+\frac{1}{24}\,{q_{{1,0}}}^{4}q_{{3,2}}+\frac{1}{8}
\,q_{{1,0}}{q_{{2,0}}}^{2}q_{{3,0}}q_{{3,1}}+{\frac {{q_{{3,0}}}^{5}
}{960}}\\
&+\frac{1}{12}\,q_{{2,1}}{q_{{2,0}}}^{3}q_{{3,0}}+\frac{1}{8}\,{q_{{3,0}}}^{2}{q
_{{2,0}}}^{2}q_{{1,1}}+\frac{1}{6}\,{q_{{1,0}}}^{3}q_{{2,0}}q_{{2,2}}+\frac{1}{6}\,{q_
{{1,0}}}^{3}q_{{3,0}}q_{{1,2}}+\frac{1}{3}\,{q_{{1,0}}}^{3}q_{{1,1}}q_{{3,1}}\\
&+
\frac{1}{4}\,{q_{{1,0}}}^{2}{q_{{2,0}}}^{2}q_{{1,2}}+\frac{1}{2}\,q_{{1,0}}{q_{{2,0}}}
^{2}{q_{{1,1}}}^{2}+\frac{1}{2}\,{q_{{1,0}}}^{2}q_{{3,0}}{q_{{1,1}}}^{2}+\frac{1}{8}\,
q_{{1,0}}q_{{2,0}}{q_{{3,0}}}^{2}q_{{2,1}}+{q_{{1,0}}}^{2}q_{{2,0}}q_{
{1,1}}q_{{2,1}} \bigg) \\
&+ \bigg( {\frac {{q_{{1,0}}}^{5}q _{{3,3}}}{120}}+{\frac
{q_{{1,1}}{q_{{3,0}}}^{5}}{320}}+{\frac {{q_{{2
,0}}}^{5}q_{{2,2}}}{240}}+\frac{1}{32}\,{q_{{2,0}}}^{4}{q_{{2,1}}}^{2}+\frac{3}{8}\,q_
{{2,0}}q_{{1,0}}{q_{{3,0}}}^{2}q_{{1,1}}q_{{2,1}}\\
&+\frac{1}{4}\,q_{{2,0}}{q_{{1
,0}}}^{2}q_{{3,0}}q_{{2,1}}q_{{3,1}}+\frac{3}{8}\,{q_{{2,0}}}^{2}q_{{1,0}}q_{{
3,0}}q_{{1,1}}q_{{3,1}}+\frac{1}{4}\,{q_{{2,0}}}^{2}q_{{1,0}}q_{{3,0}}{q_{{2,1
}}}^{2}+\frac{1}{2}\,{q_{{1,0}}}^{3}q_{{3,0}}q_{{1,1}}q_{{1,2}}\\
&+\frac{3}{4}\,{q_{{1,0}
}}^{2}{q_{{2,0}}}^{2}q_{{1,1}}q_{{1,2}}+\frac{1}{8}\,{q_{{2,0}}}^{2}q_{{1,0}}{
q_{{3,0}}}^{2}q_{{1,2}}+\frac{1}{8}\,{q_{{2,0}}}^{3}q_{{1,0}}q_{{2,1}}q_{{3,1}
}+\frac{1}{12}\,{q_{{2,0}}}^{3}q_{{1,0}}q_{{3,0}}q_{{2,2}}\\
&+\frac{1}{16}\,q_{{2,0}}{q_{
{1,0}}}^{2}{q_{{3,0}}}^{2}q_{{2,2}}+\frac{3}{2}\,q_{{2,0}}{q_{{1,0}}}^{2}{q_{{
1,1}}}^{2}q_{{2,1}}+\frac{1}{4}\,{q_{{2,0}}}^{3}q_{{3,0}}q_{{1,1}}q_{{2,1}}+\frac{1}{16}
\,{q_{{2,0}}}^{2}{q_{{1,0}}}^{2}q_{{3,0}}q_{{3,2}}\\
&+\frac{1}{2}\,{q_{{1,0}}}^
{3}q_{{2,0}}q_{{1,1}}q_{{2,2}}+\frac{1}{2}\,{q_{{1,0}}}^{3}q_{{2,0}}q_{{2,1}}q
_{{1,2}}+\frac{1}{2}\,{q_{{1,0}}}^{2}q_{{3,0}}{q_{{1,1}}}^{3}+\frac{1}{16}\,{q_{{1,0}}
}^{2}{q_{{2,0}}}^{2}{q_{{3,1}}}^{2}+\frac{1}{8}\,{q_{{1,0}}}^{4}q_{{1,1}}q_{{3
,2}}\\
&+\frac{1}{2}\,{q_{{2,0}}}^{2}q_{{1,0}}{q_{{1,1}}}^{3}+{\frac
{{q_{{2,0}}}^
{4}q_{{1,0}}q_{{3,2}}}{96}}+\frac{1}{24}\,q_{{2,0}}{q_{{1,0}}}^{4}q_{{2,3}}+{
\frac {{q_{{3,0}}}^{4}q_{{1,0}}q_{{3,1}}}{192}}+{\frac
{q_{{2,0}}{q_{{
3,0}}}^{4}q_{{2,1}}}{128}}\\
&+\frac{1}{2}\,{q_{{1,0}}}^{3}{q_{{1,1}}}^{2}q_{{3,1}
}+\frac{1}{2}\,{q_{{1,0}}}^{3}q_{{1,1}}{q_{{2,1}}}^{2}+\frac{1}{8}\,{q_{{1,0}}}^{4}q_{
{3,1}}q_{{1,2}}+\frac{1}{8}\,{q_{{1,0}}}^{4}q_{{2,1}}q_{{2,2}}+\frac{1}{12}\,{q_{{2,0}
}}^{2}{q_{{1,0}}}^{3}q_{{1,3}}\\
&+\frac{1}{32}\,{q_{{2,0}}}^{4}q_{{1,1}}q_{{3,1}}
+\frac{1}{32}\,{q_{{2,0}}}^{4}q_{{3,0}}q_{{1,2}}+{\frac
{{q_{{2,0}}}^{2}{q_{{3
,0}}}^{3}q_{{3,1}}}{96}}+\frac{1}{16}\,{q_{{1,0}}}^{2}{q_{{3,0}}}^{2}{q_{{2,1}
}}^{2}+\frac{3}{16}\,{q_{{2,0}}}^{2}{q_{{3,0}}}^{2}{q_{{1,1}}}^{2}\\
&+\frac{1}{24}\,{q_{{ 1,0}}}^{4}q_{{3,0}}q_{{1,3}} \bigg)+\mathrm{h.d.t.},
\\
\mathcal F_1 =& \frac{1}{8}\,q_{{1,1}} + \bigg({\frac
{q_{{3,0}}q_{{3,1}}}{96}}+\frac{1}{8}\,q
_{{1,0}}q_{{1,2}}+\frac{1}{16}\,{q_{{1,1}}}^{2} \bigg)
 + \bigg({\frac
{q_{{1,0}}{q_{{3,1}}}^{2}}{96}}+\frac{1}{16}\,{q_{{1,0}}}^{2}q_{{1,3}}
+{\frac
{{q_{{3,0}}}^{2}q_{{1,2}}}{96}}\\
&+\frac{1}{48}\,q_{{3,0}}{q_{{2,1}}}^{2} +{\frac
{{q_{{2,0}}}^{2}q_{{3,2}}}{64}}+\frac{1}{24}\,{q_{{1,1}}}^{3}+\frac{1}{48}\,q_
{{3,0}}q_{{1,1}}q_{{3,1}}+\frac{1}{24}\,q_{{2,0}}q_{{3,0}}q_{{2,2}}+{\frac
{q_
{{1,0}}q_{{3,0}}q_{{3,2}}}{96}}\\
&+\frac{1}{4}\,q_{{1,0}}q_{{1,1}}q_{{1,2}}+\frac{1}{24}
\,q_{{2,0}}q_{{2,1}}q_{{3,1}} \bigg) + \bigg( {\frac
{{q _{{3,0}}}^{3}q_{{3,2}}}{384}}+{\frac
{{q_{{3,0}}}^{2}{q_{{3,1}}}^{2}}{
256}}+\frac{1}{32}\,q_{{2,0}}{q_{{2,1}}}^{3}+{\frac
{{q_{{2,0}}}^{3}q_{{2,3}}
}{96}}\\
&+\frac{1}{8}\,{q_{{1,0}}}^{2}{q_{{1,2}}}^{2}+\frac{1}{48}\,{q_{{1,0}}}^{3}q_{{1,
4}}+\frac{1}{32}\,q_{{1,1}}q_{{1,0}}q_{{3,0}}q_{{3,2}}+\frac{1}{12}\,q_{{2,1}}q_{{1,0}
}q_{{3,0}}q_{{2,2}}+\frac{1}{24}\,q_{{3,0}}q_{{1,0}}q_{{3,1}}q_{{1,2}}\\
&+\frac{1}{8}\,q_
{{2,0}}q_{{1,1}}q_{{2,1}}q_{{3,1}}+\frac{1}{24}\,q_{{2,0}}q_{{1,0}}q_{{3,0}}q_
{{2,3}}+{\frac
{7\,q_{{2,0}}q_{{1,0}}q_{{2,1}}q_{{3,2}}}{96}}+\frac{1}{12}\,q_
{{2,0}}q_{{1,0}}q_{{3,1}}q_{{2,2}}\\
&+\frac{1}{8}\,q_{{2,0}}q_{{3,0}}q_{{1,1}}q_{
{2,2}}+\frac{1}{8}\,q_{{2,0}}q_{{3,0}}q_{{2,1}}q_{{1,2}}+{\frac
{{q_{{3,0}}}^{
2}q_{{1,0}}q_{{1,3}}}{96}}+\frac{1}{16}\,{q_{{2,1}}}^{2}q_{{3,0}}q_{{1,1}}+\frac{3}{8}
\,{q_{{1,1}}}^{2}q_{{1,0}}q_{{1,2}}\\
&+\frac{1}{32}\,q_{{1,1}}q_{{1,0}}{q_{{3,1}}
}^{2}+\frac{1}{32}\,q_{{1,1}}{q_{{3,0}}}^{2}q_{{1,2}}+\frac{1}{32}\,{q_{{1,1}}}^{2}q_{
{3,0}}q_{{3,1}}+\frac{1}{16}\,{q_{{2,1}}}^{2}q_{{1,0}}q_{{3,1}}+{\frac
{7\,{q_
{{2,0}}}^{2}q_{{3,0}}q_{{1,3}}}{192}}\\
&+\frac{3}{16}\,{q_{{1,0}}}^{2}q_{{1,1}}q_ {{1,3}}+{\frac
{{q_{{1,0}}}^{2}q_{{3,1}}q_{{3,2}}}{64}}+{\frac {{q_{{1
,0}}}^{2}q_{{3,0}}q_{{3,3}}}{192}}+{\frac
{{q_{{2,0}}}^{2}q_{{1,0}}q_{ {3,3}}}{64}}+{\frac
{3\,{q_{{2,0}}}^{2}q_{{1,1}}q_{{3,2}}}{64}}\\
&+{ \frac {7\,{q_{{2,0}}}^{2}q_{{2,1}}q_{{2,2}}}{96}}+{\frac
{5\,{q_{{2,0}
}}^{2}q_{{3,1}}q_{{1,2}}}{96}}+\frac{1}{32}\,{q_{{1,1}}}^{4} \bigg)
  + \mathrm{h.d.t.},
\\
\mathcal F_2 =& {\frac {3\,q_{{3,3}}}{2560}}+ \bigg(
{\frac {11\,{q_{{2 ,2}}}^{2}}{1920}}+{\frac
{9\,q_{{1,1}}q_{{3,3}}}{2560}}+{\frac {41\,q_
{{3,1}}q_{{1,3}}}{7680}}+{\frac {q_{{2,0}}q_{{2,4}}}{320}}+{\frac
{7\, q_{{2,1}}q_{{2,3}}}{960}}+{\frac
{3\,q_{{1,0}}q_{{3,4}}}{2560}}\\
&+{ \frac {49\,q_{{1,2}}q_{{3,2}}}{7680}}+{\frac
{19\,q_{{3,0}}q_{{1,4}}}{ 7680}} \bigg)  + \mathrm{h.d.t.}
\end{align*}
\end{small}

\item {\bf The $D_4$ case}\\
 The related affine Kac--Moody algebra is realized by $8\times 8$
 matrices, and
\begin{align}\label{LdrhoD4}
&\Ld_1=\sqrt{2}\left(e_{2,1}+e_{3,2}+e_{4,3}+\frac{1}{2}e_{5,3}+\frac{1}{2}e_{6,4}
+e_{6,5}+e_{7,6}+e_{8,7}+\frac{z}{2}e_{1,7}+\frac{z}{2}e_{2,8}\right), \\
&\rho=\mathrm{diag}(-3,-2,-1,0,0,1,2,3),
\end{align}
with $e_{i,j} = \Big(\delta_{i,s}\delta_{j,t}\Big)_{s,t = 1}^8.$
\begin{small}
\begin{align*}
\mathcal F_0 =& \bigg(
\frac{1}{2}\,q_{{1,0}}{q_{{2,0}}}^{2}+\frac{1}{2}\,{q_{{1,0}}}^{2}q_{{3,0}}
+\frac{1}{2}\,q_{{1,0}}{q_{{4,0}}}^{2} \bigg) + \bigg(
\frac{1}{2}\,{q_{{1,0}}}^{
2}q_{{2,0}}q_{{2,1}}+\frac{1}{2}\,{q_{{1,0}}}^{2}q_{{3,0}}q_{{1,1}}+\frac{1}{2}\,q_{{1
,0}}{q_{{2,0}}}^{2}q_{{1,1}}\\
&+\frac{1}{2}\,{q_{{1,0}}}^{2}q_{{4,0}}q_{{4,1}}+\frac{1}{2}{q_{{4,0}}}^{2}q_{{1,0}}q_{{1,1}}-\frac{1}{12}\,{q_{{4,0}}}^{2}q_{{2,0}}q_{
{3,0}}+\frac{1}{6}\,{q_{{1,0}}}^{3}q_{{3,1}}+\frac{1}{36}\,{q_{{2,0}}}^{3}q_{{3,0}}
 \bigg) \\
&+ \bigg(- \frac{1}{12}\,q_{{3,0}}q_{{2,1}}q_{{1,0}}{q_{{4
,0}}}^{2} -\frac{1}{12}\,q_{{2,0}}q_{{3,1}}q_{{1,0}}{q_{{4,0}}}^{2}-\frac{1}{6}\,{q_{{4
,0}}}^{2}q_{{2,0}}q_{{3,0}}q_{{1,1}}+\frac{1}{12}\,q_{{3,0}}q_{{2,1}}q_{{1,0}}
{q_{{2,0}}}^{2}\\
&+{q_{{1,0}}}^{2}q_{{2,0}}q_{{2,1}}q_{{1,1}}+q_{{1,1}}{q
_{{1,0}}}^{2}q_{{4,0}}q_{{4,1}}+\frac{1}{48}\,{q_{{2,0}}}^{4}q_{{2,1}}+\frac{1}{6}\,{q
_{{4,1}}}^{2}{q_{{1,0}}}^{3}+\frac{1}{24}\,{q_{{1,0}}}^{4}q_{{3,2}} -{\frac
{q_ {{2,1}}{q_{{4,0}}}^{4}}{144}}\\
&+{\frac {{q_{{4,0}}}^{2}{q_{{3,0}}}^{3}}{
216}}+\frac{1}{6}\,{q_{{1,0}}}^{3}{q_{{2,1}}}^{2}+{\frac
{{q_{{2,0}}}^{2}{q_{{
3,0}}}^{3}}{216}}+\frac{1}{6}\,q_{{4,0}}q_{{4,2}}{q_{{1,0}}}^{3}+\frac{1}{2}\,{q_{{4,0
}}}^{2}q_{{1,0}}{q_{{1,1}}}^{2}+\frac{1}{4}\,{q_{{1,0}}}^{2}q_{{1,2}}{q_{{4,0}
}}^{2}\\
&-\frac{1}{24}\,{q_{{4,0}}}^{2}{q_{{2,0}}}^{2}q_{{2,1}}-\frac{1}{18}\,{q_{{4,0}}}
^{3}q_{{4,1}}q_{{2,0}}+\frac{1}{6}\,q_{{2,0}}q_{{2,2}}{q_{{1,0}}}^{3}+\frac{1}{2}\,{q_
{{1,0}}}^{2}q_{{3,0}}{q_{{1,1}}}^{2}+\frac{1}{18}\,{q_{{2,0}}}^{3}q_{{3,0}}q_{
{1,1}}\\
&+\frac{1}{3}\,{q_{{1,0}}}^{3}q_{{3,1}}q_{{1,1}}+\frac{1}{2}\,{q_{{2,0}}}^{2}q_{{
1,0}}{q_{{1,1}}}^{2}+\frac{1}{36}\,{q_{{2,0}}}^{3}q_{{3,1}}q_{{1,0}}+\frac{1}{6}\,{q_{
{1,0}}}^{3}q_{{1,2}}q_{{3,0}}+\frac{1}{4}\,{q_{{1,0}}}^{2}q_{{1,2}}{q_{{2,0}}}
^{2}\\
&-\frac{1}{6}\,q_{{1,0}}q_{{2,0}}q_{{3,0}}q_{{4,0}}q_{{4,1}} \bigg)
 + \mathrm{h.d.t.},
\\
\mathcal F_1 =& \frac{1}{6}\,q_{{1,1}}+ \bigg(
\frac{1}{6}\,q_{{1,0}}q_{{1,2}}+\frac{1}{12}\,{q_{{1,1}} }^{2}
\bigg) + \bigg({\frac {q_{{2,0}}{q_{{2,1}}}^{2}
}{72}}+\frac{1}{12}\,{q_{{1,0}}}^{2}q_{{1,3}}-{\frac
{{q_{{4,1}}}^{2}q_{{2,0}} }{72}}+{\frac
{{q_{{3,0}}}^{2}q_{{3,1}}}{432}}\\
&+{\frac {{q_{{2,0}}}^{2} q_{{2,2}}}{72}}-{\frac
{q_{{2,2}}{q_{{4,0}}}^{2}}{72}}+\frac{1}{18}\,{q_{{1,1}
}}^{3}-\frac{1}{36}\,q_{{2,1}}q_{{4,0}}q_{{4,1}}-\frac{1}{36}\,q_{{4,0}}q_{{4,2}}q_{{2
,0}}+\frac{1}{3}\,q_{{1,0}}q_{{1,2}}q_{{1,1}} \bigg) \\
&+ \bigg( \frac{1}{24}
\,{q_{{1,1}}}^{4}-\frac{1}{36}\,q_{{1,0}}q_{{4,0}}q_{{4,3}}q_{{2,0}} -\frac{1}{18}\,q
_{{1,0}}q_{{4,1}}q_{{4,2}}q_{{2,0}}+{\frac
{q_{{3,0}}q_{{2,0}}q_{{2,1} }q_{{3,1}}}{72}}-{\frac
{q_{{4,0}}q_{{4,1}}q_{{3,0}}q_{{3,1}}}{72}}\\
&-\frac{1}{18}
\,q_{{4,0}}q_{{4,1}}q_{{1,0}}q_{{2,2}}-\frac{1}{12}\,q_{{4,0}}q_{{4,1}}q_{{2
,0}}q_{{1,2}}-\frac{1}{12}\,q_{{4,0}}q_{{4,1}}q_{{1,1}}q_{{2,1}}-\frac{1}{18}\,q_{{4,0
}}q_{{4,2}}q_{{1,0}}q_{{2,1}}\\
&-\frac{1}{12}\,q_{{4,0}}q_{{4,2}}q_{{2,0}}q_{{1,1
}}+\frac{1}{18}\,q_{{2,0}}q_{{1,0}}q_{{2,1}}q_{{2,2}}+{\frac
{q_{{3,0}}{q_{{2,0 }}}^{2}q_{{3,2}}}{216}}+{\frac
{{q_{{2,0}}}^{2}q_{{1,0}}q_{{2,3}}}{72}
}-\frac{1}{24}\,q_{{1,1}}q_{{2,2}}{q_{{4,0}}}^{2}\\
&-\frac{1}{24}\,q_{{2,1}}q_{{1,2}}{q_{ {4,0}}}^{2}-{\frac
{q_{{1,0}}q_{{2,3}}{q_{{4,0}}}^{2}}{72}}+{\frac {q_
{{4,0}}q_{{4,2}}{q_{{3,0}}}^{2}}{144}}-\frac{1}{24}\,{q_{{4,1}}}^{2}q_{{1,0}}q
_{{2,1}}-\frac{1}{24}\,{q_{{4,1}}}^{2}q_{{2,0}}q_{{1,1}}\\
&+\frac{1}{2}\,{q_{{1,1}}}^{2}q
_{{1,0}}q_{{1,2}}+\frac{1}{4}\,q_{{1,1}}{q_{{1,0}}}^{2}q_{{1,3}}+{\frac
{q_{{3
,0}}q_{{3,2}}{q_{{4,0}}}^{2}}{216}}-\frac{1}{36}\,q_{{2,0}}q_{{1,3}}{q_{{4,0}}
}^{2}+\frac{1}{24}\,q_{{1,1}}q_{{2,0}}{q_{{2,1}}}^{2}\\
&+\frac{1}{24}\,{q_{{2,0}}}^{2}q_{
{2,1}}q_{{1,2}}+\frac{1}{24}\,q_{{1,1}}{q_{{2,0}}}^{2}q_{{2,2}}+{\frac
{{q_{{3 ,0}}}^{2}q_{{2,0}}q_{{2,2}}}{144}}+{\frac
{q_{{1,1}}{q_{{3,0}}}^{2}q_{ {3,1}}}{144}}+{\frac
{{q_{{3,0}}}^{2}q_{{1,0}}q_{{3,2}}}{432}}\\
&+{\frac {q_{{3,0}}q_{{1,0}}{q_{{3,1}}}^{2}}{216}}+{\frac
{{q_{{4,1}}}^{2}{q_{{ 3,0}}}^{2}}{288}}+{\frac
{{q_{{2,0}}}^{3}q_{{1,3}}}{108}}+\frac{1}{6}\,{q_{{1,0
}}}^{2}{q_{{1,2}}}^{2}+{\frac {q_{{1,0}}{q_{{2,1}}}^{3}}{72}}+{\frac
{ {q_{{3,1}}}^{2}{q_{{4,0}}}^{2}}{432}}+{\frac
{{q_{{3,0}}}^{2}{q_{{2,1} }}^{2}}{288}}\\
&+{\frac {{q_{{3,0}}}^{3}q_{{1,2}}}{432}}+\frac{1}{36}\,q_{{1,4}}{
q_{{1,0}}}^{3}+{\frac {{q_{{2,0}}}^{2}{q_{{3,1}}}^{2}}{432}} \bigg)
 +\mathrm{h.d.t.},
\\
\mathcal F_2 =& \bigg( {\frac {q_{{3,0}}q_{{3,3}}}{1620}}+{\frac
{7\,q_{{3,1}}q_{{3,2 }}}{6480}} \bigg)  + \mathrm{h.d.t.}
\end{align*}
\end{small}
These results coincide with the ones in \cite{LRZ} up to the rescaling
\[
\ep\mapsto\sqrt{-6}\,\ep, \quad q_{\al,k}\mapsto(-1)^{k+1}3^{\dt_{\al,4}/2} q_{\al,k}.
\]

\item {\bf The $B_3$ case}\\
The related affine Kac--Moody algebra is realized by $7\times 7$
matrices.
\begin{align}
&\Ld_1=\sqrt{2}\left(e_{2,1}+e_{3,2}+e_{4,3}+e_{5,4}+e_{6,5}
+e_{7,6}+\frac{z}{2}e_{1,6}+\frac{z}{2}e_{2,7}\right), \\
&\rho=\mathrm{diag}(-3,-2,-1,0,1,2,3),
\end{align}
with $e_{i,j} = \Big(\delta_{i,s}\delta_{j,t}\Big)_{s,t = 1}^7.$
\begin{small}
\begin{align*}
\mathcal F_0 =& \bigg(
\frac{1}{2}\,q_{{1,0}}{q_{{2,0}}}^{2}+\frac{1}{2}\,{q_{{1,0}}}^{2}q_{{3,0}}
 \bigg) + \bigg(\frac{1}{36}\,{q_{{2,0}}}^{3}q_{{3,0}}+\frac{1}{6}\,{q_{{1,0
}}}^{3}q_{{3,1}}+\frac{1}{2}\,{q_{{1,0}}}^{2}q_{{2,0}}q_{{2,1}}+\frac{1}{2}\,{q_{{1,0}
}}^{2}q_{{3,0}}q_{{1,1}}\\
&+\frac{1}{2}\,q_{{1,0}}{q_{{2,0}}}^{2}q_{{1,1}}
 \bigg) + \bigg( \frac{1}{24}\,{q_{{1,0}}}^{4}q_{{3,2}}+\frac{1}{6}\,{q_
{{1,0}}}^{3}{q_{{2,1}}}^{2}+{\frac {{q_{{2,0}}}^{2}{q_{{3,0}}}^{3}}{
216}}+\frac{1}{48}\,{q_{{2,0}}}^{4}q_{{2,1}}+{q_{{1,0}}}^{2}q_{{2,0}}q_{{1,1}}
q_{{2,1}}\\
&+\frac{1}{12}\,q_{{1,0}}{q_{{2,0}}}^{2}q_{{3,0}}q_{{2,1}}+\frac{1}{36}\,q_{{1
,0}}{q_{{2,0}}}^{3}q_{{3,1}}+\frac{1}{2}\,q_{{1,0}}{q_{{2,0}}}^{2}{q_{{1,1}}}^
{2}+\frac{1}{18}\,{q_{{2,0}}}^{3}q_{{3,0}}q_{{1,1}}+\frac{1}{6}\,{q_{{1,0}}}^{3}q_{{2,0
}}q_{{2,2}}\\
&+\frac{1}{6}\,{q_{{1,0}}}^{3}q_{{3,0}}q_{{1,2}}+\frac{1}{3}\,{q_{{1,0}}}^{3
}q_{{1,1}}q_{{3,1}}+\frac{1}{4}\,{q_{{1,0}}}^{2}{q_{{2,0}}}^{2}q_{{1,2}}+\frac{1}{2}\,
{q_{{1,0}}}^{2}q_{{3,0}}{q_{{1,1}}}^{2} \bigg)  + \bigg(\frac{1}{12}\,{q_{{1,1}}}^{2}{q_{{2,0}}}^{3}q_{{3,0}}\\
&+{\frac {q_{{3,0}}{q_{{2 ,0}}}^{4}q_{{3,1}}}{288}}+{\frac
{{q_{{3,0}}}^{2}{q_{{2,0}}}^{3}q_{{2,
1}}}{72}}+\frac{1}{16}\,q_{{1,1}}{q_{{2,0}}}^{4}q_{{2,1}}+{\frac
{q_{{1,1}}{q_
{{2,0}}}^{2}{q_{{3,0}}}^{3}}{72}}+\frac{1}{12}\,{q_{{1,0}}}^{3}{q_{{2,0}}}^{2}
q_{{1,3}}\\
&+{\frac
{{q_{{1,0}}}^{2}{q_{{2,0}}}^{3}q_{{3,2}}}{72}}+\frac{1}{24}\,
{q_{{1,0}}}^{4}q_{{3,0}}q_{{1,3}}+\frac{1}{2}\,{q_{{1,0}}}^{2}q_{{3,0}}{q_{{1,
1}}}^{3}+\frac{1}{48}\,q_{{1,0}}{q_{{2,0}}}^{4}q_{{2,2}}+\frac{1}{12}\,q_{{1,0}}{q_{{2
,0}}}^{3}{q_{{2,1}}}^{2}\\
&+\frac{1}{2}\,q_{{1,0}}{q_{{2,0}}}^{2}{q_{{1,1}}}^{3}+
\frac{1}{2}\,{q_{{1,0}}}^{3}{q_{{1,1}}}^{2}q_{{3,1}}+\frac{1}{2}\,{q_{{1,0}}}^{3}q_{{1
,1}}{q_{{2,1}}}^{2}+\frac{1}{8}\,{q_{{1,0}}}^{4}q_{{1,1}}q_{{3,2}}+\frac{1}{8}\,{q_{{1
,0}}}^{4}q_{{2,1}}q_{{2,2}}\\
&+\frac{1}{8}\,{q_{{1,0}}}^{4}q_{{3,1}}q_{{1,2}}+\frac{1}{24}\,{q_{{1,0}}}^{4}q_{{2,0}}q_{{2,3}}+\frac{1}{18}\,q_{{1,0}}q_{{3,0}}{q_{{2,0
}}}^{3}q_{{1,2}}+{\frac
{q_{{1,0}}{q_{{3,0}}}^{2}{q_{{2,0}}}^{2}q_{{3, 1}}}{72}}+{\frac
{q_{{1,0}}{q_{{3,0}}}^{3}q_{{2,0}}q_{{2,1}}}{108}}\\
&+\frac{3}{2}\,{q_{{1,0}}}^{2}q_{{2,0}}{q_{{1,1}}}^{2}q_{{2,1}}+\frac{3}{4}\,{q_{{1,0}}}^{
2}{q_{{2,0}}}^{2}q_{{1,1}}q_{{1,2}}+\frac{1}{12}\,{q_{{1,0}}}^{2}{q_{{2,0}}}^{
2}q_{{2,1}}q_{{3,1}}+\frac{1}{2}\,{q_{{1,0}}}^{3}q_{{3,0}}q_{{1,1}}q_{{1,2}}\\
&+\frac{1}{24}\,{q_{{1,0}}}^{2}q_{{3,0}}{q_{{2,0}}}^{2}q_{{2,2}}+\frac{1}{12}\,{q_{{1,0}}
}^{2}q_{{3,0}}q_{{2,0}}{q_{{2,1}}}^{2}+\frac{1}{2}\,{q_{{1,0}}}^{3}q_{{2,0}}q_
{{1,1}}q_{{2,2}}+\frac{1}{2}\,{q_{{1,0}}}^{3}q_{{2,0}}q_{{2,1}}q_{{1,2}}\\
&+\frac{1}{12} \,q_{{1,0}}{q_{{2,0}}}^{3}q_{{1,1}}q_{{3,1}}+{\frac
{{q_{{1,0}}}^{5}q_ {{3,3}}}{120}}+{\frac
{{q_{{2,0}}}^{5}q_{{1,2}}}{144}}+\frac{1}{4}\,q_{{1,0}}q
_{{3,0}}{q_{{2,0}}}^{2}q_{{1,1}}q_{{2,1}} \bigg)
+\mathrm{h.d.t.},
\end{align*}
\end{small}
\begin{small}
\begin{align*}
\mathcal F_1 =& \frac{1}{6}\,q_{{1,1}} + \bigg(
\frac{1}{6}\,q_{{1,0}}q_{{1,2}}+\frac{1}{12}\,{q_{{1,1}} }^{2}
\bigg)  + \bigg(
\frac{1}{12}\,{q_{{1,0}}}^{2}q_{{1,3}}+{ \frac
{q_{{2,0}}{q_{{2,1}}}^{2}}{72}}+{\frac {{q_{{2,0}}}^{2}q_{{2,2}}
}{72}}+{\frac
{{q_{{3,0}}}^{2}q_{{3,1}}}{432}}\\
&+\frac{1}{18}\,{q_{{1,1}}}^{3}+1 /3\,q_{{1,0}}q_{{1,1}}q_{{1,2}}
\bigg)  + \bigg( {\frac
{{q_{{2,0}}}^{3}q_{{1,3}}}{108}}+{\frac
{{q_{{2,0}}}^{2}{q_{{3,1}}}^{2 }}{432}}+{\frac
{{q_{{3,0}}}^{3}q_{{1,2}}}{432}}+{\frac {q_{{1,0}}{q_{
{2,1}}}^{3}}{72}}\\
&+{\frac {{q_{{3,0}}}^{2}{q_{{2,1}}}^{2}}{288}}+\frac{1}{6}\,{
q_{{1,0}}}^{2}{q_{{1,2}}}^{2}+\frac{1}{36}\,{q_{{1,0}}}^{3}q_{{1,4}}+\frac{1}{18}\,q_{
{1,0}}q_{{2,0}}q_{{2,1}}q_{{2,2}}+{\frac
{q_{{3,0}}q_{{2,0}}q_{{2,1}}q
_{{3,1}}}{72}}+\frac{1}{24}\,{q_{{1,1}}}^{4}\\
&+\frac{1}{24}\,q_{{1,1}}q_{{2,0}}{q_{{2,1} }}^{2}+{\frac
{q_{{3,0}}{q_{{2,0}}}^{2}q_{{3,2}}}{216}}+{\frac {{q_{{3
,0}}}^{2}q_{{2,0}}q_{{2,2}}}{144}}+\frac{1}{24}\,q_{{1,1}}{q_{{2,0}}}^{2}q_{{2
,2}}+{\frac
{{q_{{3,0}}}^{2}q_{{1,1}}q_{{3,1}}}{144}}\\
&+\frac{1}{24}\,{q_{{2,0}} }^{2}q_{{2,1}}q_{{1,2}}+{\frac
{q_{{1,0}}{q_{{2,0}}}^{2}q_{{2,3}}}{72}
}+\frac{1}{4}\,{q_{{1,0}}}^{2}q_{{1,1}}q_{{1,3}}+\frac{1}{2}\,q_{{1,0}}{q_{{1,1}}}^{2}
q_{{1,2}}+{\frac {q_{{1,0}}{q_{{3,0}}}^{2}q_{{3,2}}}{432}}\\
&+{\frac {q_{ {1,0}}q_{{3,0}}{q_{{3,1}}}^{2}}{216}} \bigg)
 +\mathrm{h.d.t.},
\\
\mathcal F_2 =& \bigg( {\frac {7\,q_{{3,1}}q_{{3,2}}}{6480}}+{\frac
{q_{{3,0}}q_{{3,3 }}}{1620}} \bigg)   +\mathrm{h.d.t.}
\end{align*}
\end{small}

One can check directly that this case can be deduced from the one of
$D_4$ by putting $q_{4,k} = 0$, as expected by Theorem
\ref{reductthm}. We also verified, for the first terms of the
expansion, the reduction from $B_3$ to $G_2$ by putting $q_{2,k}=0$.

\item {\bf The $C_2$ case}\\
The elements $\Ld_1$ and $\rho$ in the reduced string equation are
given by \eqref{LdrhoA3}.
\begin{small}
\begin{align*}
\mathcal F_0 =& \frac{1}{2}\,{q_{{1,0}}}^{2}q_{{2,0}} + \bigg(
\frac{1}{6}\,{q_{{1,0}}}^{3}q_{{3
,1}}+\frac{1}{2}\,{q_{{1,0}}}^{2}q_{{1,1}}q_{{2,0}} \bigg)
 +
 \bigg( \frac{1}{24}\,q_{{2,2}}{q_{{1,0}}}^{4}+\frac{1}{3}\,q_{{1,1}}{q_{{1,0}}}^{3}q_
{{3,1}}\\
&+\frac{1}{2}\,{q_{{1,1}}}^{2}q_{{2,0}}{q_{{1,0}}}^{2}+{\frac
{{q_{{2,0}
}}^{5}}{960}}+\frac{1}{6}\,q_{{1,2}}{q_{{1,0}}}^{3}q_{{2,0}} \bigg)
 + \bigg(
\frac{1}{2}\,q_{{1,1}}{q_{{1,0}}}^{3}q_{{1,2}}q_{{2,0}}+{\frac {{
q_{{2,0}}}^{5}q_{{1,1}}}{320}}\\
&+{\frac {q_{{2,3}}{q_{{1,0}}}^{5}}{120}}
+\frac{1}{8}\,q_{{2,2}}q_{{1,1}}{q_{{1,0}}}^{4}+\frac{1}{8}\,q_{{1,2}}q_{{2,1}}{q_{{1,0
}}}^{4}+\frac{1}{24}\,q_{{1,3}}q_{{2,0}}{q_{{1,0}}}^{4}+{\frac
{{q_{{2,0}}}^{4
}q_{{1,0}}q_{{2,1}}}{192}}\\
&+\frac{1}{2}\,q_{{2,1}}{q_{{1,1}}}^{2}{q_{{1,0}}}^{3
}+\frac{1}{2}\,{q_{{1,1}}}^{3}q_{{2,0}}{q_{{1,0}}}^{2} \bigg)
  + \mathrm{h.d.t.},
\\
\mathcal F_1 =& \frac{1}{8}\,q_{{1,1}} + \bigg(
\frac{1}{16}\,{q_{{1,1}}}^{2}+\frac{1}{8}\,q_{{1,0}}q_{{
1,2}}+{\frac {q_{{2,1}}q_{{2,0}}}{96}} \bigg)  + \bigg(
\frac{1}{4}\,q_{{1,2}}q_{{1,1}}q_{{1,0}}+\frac{1}{48}\,q_{{1,1}}q_{{2,0}}q_{{2,1}}\\
&+{ \frac
{q_{{2,2}}q_{{2,0}}q_{{1,0}}}{96}}+\frac{1}{24}\,{q_{{1,1}}}^{3}+{\frac
{{q_{{2,1}}}^{2}q_{{1,0}}}{96}}+{\frac
{q_{{1,2}}{q_{{2,0}}}^{2}}{96}}
+\frac{1}{16}\,q_{{1,3}}{q_{{1,0}}}^{2} \bigg)  + \bigg(
+\frac{1}{32}\,q _{{1,2}}q_{{1,1}}{q_{{2,0}}}^{2}\\
&+{\frac {q_{{1,3}}{q_{{2,0}}}^{2}q_{{1
,0}}}{96}}+\frac{1}{32}\,{q_{{1,1}}}^{2}q_{{2,0}}q_{{2,1}}+\frac{1}{32}\,{q_{{2,1}}}^{
2}q_{{1,1}}q_{{1,0}}+\frac{3}{16}\,q_{{1,3}}q_{{1,1}}{q_{{1,0}}}^{2}+{\frac
{q _{{3,2}}q_{{2,1}}{q_{{1,0}}}^{2}}{64}}\\
&+{\frac {q_{{2,3}}q_{{2,0}}{q_{{
1,0}}}^{2}}{192}}+\frac{3}{8}\,{q_{{1,1}}}^{2}q_{{1,0}}q_{{1,2}}+\frac{1}{32}\,{q_{{1,
1}}}^{4}+\frac{1}{24}\,q_{{1,2}}q_{{2,1}}q_{{1,0}}q_{{2,0}}+\frac{1}{32}\,q_{{2,2}}q_{
{1,1}}q_{{2,0}}q_{{1,0}}\\
&+\frac{1}{8}\,{q_{{1,2}}}^{2}{q_{{1,0}}}^{2}+\frac{1}{48}\,q_{
{1,4}}{q_{{1,0}}}^{3}+{\frac {q_{{2,2}}{q_{{2,0}}}^{3}}{384}}+{\frac
{ {q_{{2,1}}}^{2}{q_{{2,0}}}^{2}}{256}} \bigg)  +
\mathrm{h.d.t.},
\\
\mathcal F_2 =&{\frac {3\,q_{{2,3}} }{2560}}+ \bigg(
{\frac {9\,q_{{2,3 }}q_{{1,1}}}{2560}}+{\frac
{41\,q_{{1,3}}q_{{2,1}}}{7680}}+{\frac {49
\,q_{{2,2}}q_{{1,2}}}{7680}}+{\frac {3\,q_{{2,4}}q_{{1,0}}}{2560}}+{
\frac {19\,q_{{1,4}}q_{{2,0}}}{7680}} \bigg)   +
\mathrm{h.d.t.}.
\end{align*}
\end{small}

One can check directly that the results can be deduced from that of
the $A_3$ case by putting $q_{2,k} = 0$ and redenoting $q_{3,k}$ as
$q_{2,k}$, as expected by Theorem \ref{reductthm}.
\end{itemize}

%\textcolor{magenta}{
%\begin{rmk}
%The method also works for Drinfeld--Sokolov hierarchies of types $E_6$ (see \cite{HKW}), and the result coincides with that in \cite{FJR}. By using theorem~\ref{reductthm}, the topological solution of the case $F_4$ can be also obtained.
%\end{rmk}
%}
\begin{rmk}
Recently Bertola, Dubrovin and Yang \cite{BDY} obtained generating
functions different from \eqref{freeenergy} for multi-point
correlation functions
$\langle\tau_{k_1}\tau_{k_2}\dots\tau_{k_n}\rangle_g$ of the KdV
hierarchy. In an alternative way, we can derive similar generating
functions for the cases $A_{\ell}$, $B_{\ell}$, $C_{\ell}$,
$D_{\ell}$ and $G_2$, by using the solutions of the reduced string
equation for the Drinfeld--Sokolov hierarchies, see [arXiv: 1505.00556]. This further
application of the reduced string equation, will be published in
another occasion.
\end{rmk}

\vskip 0.5truecm \noindent{\bf Acknowledgments.} The author M.\,C.
thanks the School of Mathematics and Computational Science of the
Sun Yat-Sen University for the warm hospitality during the
preparation of this article; C.-Z.\,W. thanks Youjin Zhang, Si-Qi
Liu, Di Yang and Jianxun Hu for helpful discussions. This work is
partially supported by the program ARIANES of the University of
Angers, the French research project DIADEMS by ANR and the region
Pays de la Loire through a project ``Nouvelle \'Equipe" (to M.\,C.),
as well as by the National Natural Science Foundation of China
No.11771461 and No.11831017 (to  C.-Z.\,W.). M.C. was supported by the European Union Horizon 2020 research and innovation program under the
Marie Sklodowska-Curie RISE 2017 grant agreement no. 778010 IPaDEGAN.

\begin{appendices}

\section{Matrix realization of affine Kac--Moody algebras}
\label{sec-app2}

For the convenience of the readers, let us report the matrix
realization of the affine Kac--Moody algebras used in the previous
section, see \cite{DS, Kac}. For each case, we report the following
data:
\begin{itemize}
    \item The size $n$ of the matrix realization, and $e_{i,j}$ means the $n\times n$ matrix whose
    $(i,j)$-component is $1$ while the others vanish.
    \item The generators $E_i,F_i$ and $H_i$ of $\mathring\fg$, and the normalization
    constant     $\kappa$ such that $(X |Y)_0=\kappa\mathrm{Tr}(X Y)$ gives the
    standard invariant bilinear form.
    \item The Cartan matrix $A$ and the Kac labels $\bf k$ of the affine Kac--Moody algebra $\fg$.
    \item The Coxeter number $h$, the set of exponents $\mathscr{E}$ of $\fg$ and the set of
    generators $\{\Lambda_j |\; j \in \mathscr{E}\}$ of its principal Heisenberg subalgebra (these
    generators are normalized as in \cite{LWZ, WuVir}).
\end{itemize}
\subsection*{Type $A_{\ell}^{(1)}$, $\ell\ge1$}
    \begin{itemize}
    \item $n = \ell+1$.
    \item $E_0=e_{1,n},  \quad F_0=e_{n,1}, \quad H_0=e_{1,1}-e_{n,n};$ \\
 $E_i=e_{i+1,i},\quad F_i=e_{i,i+1}, \quad H_i=-e_{i,i}+e_{i+1,i+1}
~~~(1\le i\le \ell).$\\
    $\kappa=1$. %, \quad  $(X\mid Y)_0=\mathrm{Tr}(XY)$.
    \item $
A=\left(
    \begin{array}{cccccc}
      2  & -1  & 0  & \cdots  &  0 & -1  \\
      -1  & 2  &  -1 &  \ddots &   &  0 \\
       0 & -1  & 2  & \ddots  &  \ddots & \vdots  \\
       \vdots & \ddots  & \ddots  & \ddots  & \ddots  & 0  \\
       0 &   & \ddots  &  \ddots &  2 &  -1 \\
       -1 &  0 & \cdots  & 0  & -1  &  2 \\
    \end{array}
  \right)_{(\ell+1)\times(\ell+1)}$, \\
except $A=\left(
          \begin{array}{cc}
            2 & -2 \\
           -2 & 2 \\
          \end{array}
        \right)$ for $\ell=1$.\\
$\bf k$ = $(k_0, k_1, \dots, k_\ell)=(1,1,\dots,1)$.
    \item $h=\ell+1, \quad \mathscr{E}=\Z\setminus h\Z, \quad
\{\Ld_j=\Ld^j\mid  j\in \mathscr{E}\}.$
\end{itemize}

\subsection*{Type $B_\ell^{(1)}$, $\ell\ge3$}
\begin{itemize}
    \item $n =2 \ell+1$.
    \item $E_0=\frac{1}{2}(e_{1,2\ell}+e_{2, 2\ell+1}), \quad
F_0=2(e_{2\ell,1}+e_{2\ell+1,2}), \quad
H_0=e_{1,1}+e_{2,2}-e_{2\ell,2\ell}-e_{2\ell+1,2\ell+1}$; \\
$E_i=e_{i+1,i}+e_{2\ell+2-i,2\ell+1-i}, \quad F_i=e_{i,i+1}+e_{2\ell+1-i,2\ell+2-i}$, \\
$H_i=-e_{i,i}+e_{i+1,i+1}-e_{2\ell+1-i,2\ell+1-i}+e_{2\ell+2-i,2\ell+2-i}
~~~(1\le i\le
\ell-1)$; \\
$E_\ell=e_{\ell+1,\ell}+e_{\ell+2,\ell+1}, \quad
F_\ell=e_{\ell,\ell+1}+e_{\ell+1,\ell+2},\quad
H_\ell=-e_{\ell,\ell}+e_{\ell+2,\ell+2}.$\\
 $\kappa=\dfrac{1}{2}$. %    $(X\mid Y)_0=\dfrac{1}{2}\mathrm{Tr}(X Y)$.
    \item $
A=\left(
    \begin{array}{cccccc}
      2  & 0  & -1  & 0 & \cdots   & 0  \\
      0  & 2  &  -1 &   &   &  \vdots \\
      -1 & -1  & 2  & \ddots  &   & 0  \\
       0 &  & \ddots  & \ddots  & -1  & 0  \\
      \vdots &   &   &  -1 &  2 &  -1 \\
      0 &  \cdots &  0 & 0  & -2  &  2 \\
    \end{array}
  \right)_{(\ell+1)\times(\ell+1)},
$\\
${\bf k} = (k_0, k_1, k_2, \dots, k_\ell)=(1,1,2,\dots,2)$.
\item $h=2 \ell, \quad \mathscr{E}=\Z^{\mathrm{odd}}$, \quad
$\{\Ld_k=\sqrt{2}\,\Ld^k, \;
\Ld_{-k}=\sqrt{2}(z^{-1}\Ld^{2\ell-1})^{k} \mid
k\in\Z^{\mathrm{odd}}_+\}$.
\end{itemize}

\subsection*{Type $C_\ell^{(1)}$, $\ell\ge2$}
\begin{itemize}
    \item $n =2 \ell$.
    \item $E_0=e_{1,2 \ell},  \quad F_0=e_{2 \ell,1}, \quad
    H_0=e_{1,1}-e_{2\ell,2\ell}$; \\
$E_i=e_{i+1,i}+e_{2\ell+1-i,2\ell-i},\quad F_i=e_{i,i+1}+e_{2\ell-i,2\ell+1-i}$, \\
$H_i=-e_{i,i}+e_{i+1,i+1}-e_{2\ell-i,2\ell-i}+e_{2\ell+1-i,2\ell+1-i}
~~~(1\le i\le \ell-1)$; \\
$E_\ell=e_{\ell+1,\ell}, \quad F_\ell=e_{\ell,\ell+1}, \quad H_\ell=-e_{\ell,\ell}+e_{\ell+1,\ell+1}.$\\
    $\kappa=1$. % $(X\mid Y)_0=\mathrm{Tr}(X Y)$.
    \item $
A=\left(
    \begin{array}{cccccc}
      2  & -1  & 0  & \cdots  &  & 0  \\
      -2  & 2  &  -1 &  \ddots &   &   \\
       0 & -1  & 2  & \ddots  &  \ddots & \vdots  \\
       \vdots & \ddots  & \ddots  & \ddots  & -1 & 0  \\
        &   & \ddots  &  \ddots &  2 &  -2 \\
      0 &   & \cdots  & 0  & -1  &  2 \\
    \end{array}
  \right)_{(\ell+1)\times(\ell+1)}$,\\
${\bf k} = (k_0, k_1, \dots, k_{\ell-1}, k_\ell)=(1,2,\dots,2,1)$.
\item $h=2 \ell, \quad \mathscr{E}=\Z^{\mathrm{odd}}$, \quad
$\{\Ld_k=\Ld^k \mid k\in\Z^{\mathrm{odd}}\}$.
\end{itemize}

\subsection*{Type $D_\ell^{(1)}$, $\ell\ge4$}
\begin{itemize}
    \item $n =2 \ell$.
    \item $E_0=\frac{1}{2}(e_{1,2\ell-1}+e_{2, 2\ell}), \quad
F_0=2(e_{2\ell-1,1}+e_{2\ell,2}),\quad
H_0=e_{1,1}+e_{2,2}-e_{2\ell-1,2\ell-1}-e_{2\ell,2\ell}$; \\
$E_i=e_{i+1,i}+e_{2\ell+1-i,2\ell-i}, \quad F_i=e_{i,i+1}+e_{2\ell-i,2\ell+1-i}$, \\
$H_i=-e_{i,i}+e_{i+1,i+1}-e_{2\ell-i,2\ell-i}+e_{2\ell+1-i,2\ell+1-i}
~~~(1\le i\le
\ell-1)$; \\
$E_\ell=\frac{1}{2}(e_{\ell+1,\ell-1}+e_{\ell+2,\ell}), \quad F_\ell={2}(e_{\ell-1,\ell+1}+e_{\ell,\ell+2})$,\\
$H_\ell=-e_{\ell-1,\ell-1}-e_{\ell,\ell}+e_{\ell+1,\ell+1}+e_{\ell+2,\ell+2}$. \\
  $\kappa=\dfrac{1}{2}$. %    $(X\mid Y)_0=\frac{1}{2}\mathrm{Tr}(X Y)$.
    \item $
A=\left(
    \begin{array}{cccccc}
      2  & 0  & -1  & 0 & \cdots   & 0  \\
      0  & 2  &  -1 &   &   &  \vdots \\
      -1 & -1  & 2  & \ddots  &   & 0  \\
       0 &  & \ddots  & \ddots  & -1  & -1  \\
      \vdots &   &   &  -1 &  2 &  0 \\
      0 &  \cdots &  0 & -1  & 0  &  2 \\
    \end{array}
  \right)_{(\ell+1)\times(\ell+1)}$,\\
${\bf k} = (k_0, k_1, \dots, k_\ell)=(1,1,2,\dots,2,1,1)$.
\item $h=2 \ell-2, \quad \mathscr{E}=\{1, 3, 5,\ldots,2\ell-3\}\cup\{(\ell-1)'\}+h\Z$
(the notation $(\ell-1)'$ means that when $\ell$ is even then each
exponent congruent to $\ell-1$ modulo $h$ has multiplicity $2$),
\\
$\{\Ld_k=\sqrt{2}\,\Ld^k, \,
\Ld_{k(\ell-1)'}=\sqrt{2\ell-2}\,\Gm^k\mid k\in\Z^{\mathrm{odd}}\}$
in
 which
\[
\Gm=-\chi\Big(
e_{\ell,1}-\frac{1}{2}e_{\ell+1,1}-\frac{z}{2}e_{\ell,2\ell}+\frac{z}{4}e_{\ell+1,2\ell}+
(-1)^\ell\big(e_{2\ell,\ell+1}-\frac{1}{2}e_{2\ell,\ell}-\frac{z}{2}e_{1,\ell+1}+\frac{z}{4}e_{1,\ell}\big)\Big)
\label{Gm}
\]
with~$\chi=1$ when~$\ell$ is even and~$\chi=\sqrt{-1}$ when~$\ell$ is
odd, and $\Ld^{j}=(z^{-1}\Ld^{2\ell-3})^{-j}$,
$\Gm^{j}=(z^{-1}\Gm)^{-j}$ for $j<0$.
\end{itemize}

\subsection*{Type $G_2^{(1)}$}
\begin{itemize}
    \item $n =7$.
    \item $E_0=\frac{1}{2}(e_{1,6}+e_{2, 7}), \quad
F_0=2(e_{6,1}+e_{7,2})$, \\
$E_1=e_{2,1}+e_{4,3}+e_{5,4}+e_{7,6}, \quad  F_1=e_{1,2}+2 e_{3,4}+2 e_{4,5}+e_{6,7}$, \\
$E_2=e_{3,2}+e_{6,5}, \quad F_2=e_{2,3}+e_{5,6}$,\\
$H_i=[E_i, F_i] \quad (i=0,1,2).$\\
 $\kappa=\dfrac{1}{2}$. %    $(X\mid Y)_0=\dfrac{1}{2}\mathrm{Tr}(X Y)$.
    \item $
A=\left(
    \begin{array}{ccc}
      2  & 0  & -1   \\
      0  & 2  &  -3 \\
      -1 & -1  & 2
    \end{array}
  \right)$, \quad
${\bf k} = (k_0, k_1, k_2)=(1,3,2)$.
\item $h=6, \quad \mathscr{E}=\{1, 5\}+6\Z$, \quad
$\{\Ld_k=\sqrt{2}\,\Ld^k, \; \Ld_{-k}=\sqrt{2}(z^{-1}\Ld^{5})^{k}
\mid k\in \mathscr{E}_{>0}\}$.
\end{itemize}

%\newpage

\end{appendices}

%\bibliographystyle{plain}
%\bibliography{/Users/mattiacafasso/Documents/BibDeskLibrary.bib}
\def\cprime{$'$}

\end{document}